\documentstyle{mn}

\def\today{\ifcase\month\or
 January\or February\or March\or April\or May\or June\or
 July\or August\or September\or October\or November\or
 December\fi\space\number\day, \number\year}

\def\todmy{\number\day\space\ifcase\month\or
 January\or February\or March\or April\or May\o r June\or
 July\or August\or September\or October\or November\or
 December\fi\space\number\year}

\def\grape{{\small GRAPE }}
\def\etal{{\it et al.} }

\title{Central cusp due to a super-massive black hole in axisymmetric
models of elliptical galaxies}
\author[F. Leeuwin, E. Athanassoula]
       { F. Leeuwin \&
       E. Athanassoula \\
 Observatoire de Marseille, 2 Place Le Verrier, 
F-13248 Marseille Cedex 4, France \\}

\date{Accepted .
      Received ;
      }

\pagerange{\pageref{firstpage}--\pageref{lastpage}}
\pubyear{2000}

\begin{document}

\maketitle
\label{firstpage}
\begin{abstract}

We use numerical simulations to investigate the cusp at
the centre of elliptical galaxies, due to
the slow growth of a super-massive black hole. We study
this problem for axisymmetric models of galaxies, with or without rotation.
 The numerical simulations are based on
the `Perturbation Particles' method, and use GRAPEs to compute
the force due to the cusp. We study how the density cusp is affected by
the initial flattening of the model, as well as the role played by
initial rotation. The logarithmic slope of the density cusp is found
to be very much insensitive to flattening; as a consequence, 
we deduce that tangential velocity anisotropy-- which supports the
flattening-- is also of little influence on the final cusp. We
investigate {\it via} two different kinds of rotating models the
efficiency with which a rotation velocity component builds within
the cusp. A cusp in rotation develops only for models where a net 
rotation component is initially present
at high energy levels. The eventual
observation of a central rotational velocity peak in E galaxies has
therefore some implications for the galaxy dynamical history.

\end{abstract}

\begin{keywords}
galaxies: structure -- galaxies: dynamics -- methods:
numerical-- galaxies: nuclei
\end{keywords}

\section{Introduction}
\label{sec_intro}
\indent

HST observations have shown that
a central density cusp is present in most, if not all, elliptical
galaxies (Lauer {\it et al.} 1995, and companion papers).
Furthermore the slope of the density cusp has shown a dichotomy
between  high mass (luminosity) systems and low mass (luminosity)
ones, analogous to what has been found for other observational
properties (Bender {\it et al.} 1989).
Low mass ellipticals tend to host steep cusps, 
the radial profile of their
luminosity $\rho(r)$ having a logarithmic slope 
 $\gamma\equiv - {\rm d } log(\rho)/{\rm d} log(r)  \simeq 1.9
\pm 0.5 $. 
On the other hand, luminous ellipticals have shallower cusps, with 
$\gamma \simeq 0.8 \pm 0.5$ (Gebhard \etal 1996).

The most popular explanation for the formation of a density cusp
is a central super-massive black hole (BH)-- see 
Richstone {\it et al.} (1998), Kormendy \& Richstone (1995)
 for reviews. Such BHs are believed to be present in a large
 fraction (possibly close to 
1, Haenelt \& Rees 1993, Tremaine 1997) of present day galaxies.
Their influence on a collisionless galactic nucleus was
first addressed by Peebles (1972), Young (1980), Goodman \& Binney
(1984) using semi-analytic models.

All these models suppose an isolated galaxy, at the center of which a
 BH grows by gas accretion. The slope of the cusp produced in such
 models is within the range observed for the less massive E galaxies.
In this paper we shall also consider the formation of a cusp in an
isolated galaxy, as we are interested in the origin for the
steep cusp ($\gamma \simeq 1.9$) of low luminosity ellipticals. 
On the other hand, numerical simulations (Makino \& Ebisuzaki, 1996,
 Quinlan \& Hernquist
 1997, Nakano \& Makino 1999) have shown that
a BH sinking within the core of an elliptical galaxy, or
a binary of BHs produce a shallow cusp, similar to
that observed for massive E galaxies. Such a BH binary could result 
from a merger. This picture is therefore 
compatible with the general belief (e.g. Nieto \& Bender 1989)
that the apparent existence of two classes of ellipticals is related to 
a more important role played by merging and interactions 
in the history of massive E galaxies.

All the semi-analytic models of cusp formation due to a central BH
 use the adiabatic invariance of actions to derive
the distribution function after a BH has slowly grown (see Young
 1980). Since actions
are explicitly known for spherical systems, analytical models
have been derived in spherical symmetry. Based on such quasi-analytic
computations for spherical models, Quinlan \etal (1995)
investigated the consequences
 of the initial density profile on the slope of the cusp, and addressed
the influence of a velocity anisotropy.

For non-spherical systems -- except for St\"ackel 
potentials -- an expression for 
the actions is unknown, and therefore models based on the conservation
of actions have not been computed hitherto.
 In this paper, we propose to use numerical simulations
 with high central resolution,
based on the PP method (Leeuwin \etal 1993), to
investigate the cusp formed in axisymmetric systems by the growth of a
BH. 
 
Many low luminosity galaxies display little evidence for triaxiality,
but are close to axisymmetry; they   
can be modeled as two-integral models, with $f(E,L_z)$.
Such a distribution function (d.f.)
 implies velocity dispersions obeying
$\sigma_R=\sigma_z$, meaning that flattening would be
 due to an excess of azimuthal motion
 (which increases the net rotation, or the tangential anisotropy).  
 This does not preclude that three-integral
models would not do as well, or better ({\it e.g.} review by Merritt, 1998).
 Many of those galaxies however exhibit kinematic features
consistent with isotropic rotators models: at least for them,
an excess of tangential motion is plausibly the main support
for their flattening.

The BH itself may have perturbed orbits sufficiently to erase
the part of memory for initial conditions that corresponds
 to conservation of a third integral (Norman \etal 1985, Gerhard
\& Binney 1985, Merritt \& Quinlan 1998).
In this case axisymmetry is a consequence of the BH growth.
Nevertheless, one still ought to investigate
the scenario in which, for some of these galaxies, the
presently observed axisymmetry already existed before the BH growth.

This is the goal of this paper, where we will study the cusp generated
by the growth of a central BH within axisymmetric systems having various
degrees of rotation.
If a BH grows in a roughly axisymmetric galaxy,  
do the initial flattening and rotation
have a sizeable effect on the evolution,
so that it deviates from that
of the well known spherical case?
Could we in certain cases be able, eventually with more detailed
observational data, 
to infer from the properties of presently observed cusps any
information about the pre-black hole galaxy? 
Should one be cautious in certain cases about deriving the BH mass
using the spherical adiabatic model?

This paper is organized as follows: first, the existing models
are briefly recalled in section 2. We then specify the initial
conditions (section 3) and the numerical
techniques used for this work (section 4).
 Results for non-rotating and rotating models are given
in sections 5 and 6, respectively. We conclude in section 7.

\section{Previous models for galactic cusps}
\label{sec:models}

Previous models are based on a few basic assumptions, and ours will
follow suit. The growth of the BH is supposed to be due to 
 gas accretion, and thus does not deplete the original stellar
component. The existence of the gaseous component is
otherwise neglected, so that the models follow 
the evolution of a purely stellar component. 
The growth time $t_{BH}$ of the black hole is taken  
to be at maximum $\sim 10^7$ Yrs. This is
long enough with respect to the stellar dynamical times, so that we can compare
our models with the adiabatic computations, but not too long for obvious
computation time reasons. It is furthermore reasonnable if
compared to the `Salpeter time', or timescale for the mass
accretion onto a black hole:
\begin{equation}
t_S = M_{BH}/\dot{M}_{BH}\simeq \epsilon \times 4 \times 10^8 Yrs .
\end{equation}
In the above, $\epsilon$ is the radiative efficiency, which is most
often $\epsilon \leq 0.1$ in current models of accretion disks
({\it cf.} Richstone {\it et al.} 1998), or in estimates derived from
quasars number counts (van der Marel 1997).
   
Furthermore the models neglect 
collisions. It has been estimated in the past that
only some of the spiral galaxies bulges are dense enough to
experience significant two-body interactions (see
{\it e.g.}  Kormendy 1988), although these estimates
were based on the belief that elliptical galaxies had cores.
Let us suppose for instance that up to 
$\sim 0.1 -1\%$ of an E galaxy mass,
corresponding to say $N \sim 10^6-10^9$ stars, is enclosed
in the central $\sim 10 $ pc, where velocity is of order a few times
$\sim 100 $ km/s. The crossing time in these regions should be
$T_{cr} \simeq 10^5 \; \left(\frac {R( {\rm pc})}{10 } \right)
\; \left(\frac{100}{ \sigma ({\rm km/s}) }\right )$ Yrs.
 Therefore the relaxation 
time would be ({\it e.g.} Binney \& Tremaine 1987):
$t_{rel} \simeq N/(8 {\rm ln} N) \, t_{cr} \simeq 10^{9}  \;
 \left(\frac {R( {\rm pc} )}{10 } \right)
\; \left(\frac{100}{ \sigma ({\rm km/s}) }\right) \;
 \left(\frac{N}{10^6}\right) $
 Yrs. Stellar collisions are liable to play some role over
the age of the universe in such a system with an already 
highly concentrated density.
For the models we consider here, however, a collisionless model is
 justified, given that the initial
galaxy has a large core, and that the BH grows in $\sim 10^7$ Yrs.
Nevertheless it should be kept in mind that collisional processes 
could play a role in the evolution of the center-most part after the
cusp has developed.

The spatial extent of the cusp is given approximately by
the {\it influence radius} $r_I$ of the BH, which is the distance up to
which the BH potential dominates. This is estimated as:
\begin{equation}
 \frac{M_{BH}}{r_I} \simeq  \sigma^2_0,
\label{eq:ri}
\end{equation} 
where $\sigma_0$ is the central velocity dispersion, and $M_{BH}$ is
the BH mass.

\subsection{Analytical models: spherical adiabatic models}
\label{ss_analytic}

Analytical models assume that the central BH grows
over time scales that are long compared
to dynamical times in the central region.
The final d.f. can then be deduced from adiabatic
 conservation of actions (Peebles 1972, Young 1980).

 In an isotropic system having initially a core, the cusp is predicted to have
a slope $\gamma = 3/2$ (Young 1980; see also Tremaine 1997).
This value, however, holds if the BH mass is less
than roughly the core mass; when larger, the self-gravity
of the cusp becomes important and the slope increases up
 to values $\simeq 2$ (e.g. Young 1980).
 For centres that are initially `non-analytical'
-- {\it i.e.} whose density profile has a non-zero derivative at the origin--
 Quinlan \etal (1995) derive larger slopes $\gamma$, 
with higher values for stronger initial cusps.

Goodman \& Binney (1984) provide the following simple picture for
the way orbits are transformed as the cusp grows.
 The fact that radial action is conserved plays an especially 
important role in the shape evolution of 
elongated orbits. It means that the radial
excursion $r_{max}- r_{min}$ does not change much, but the potential center,
which is originally at the orbit center (roughly harmonic potential),
 is finally at one of its foci (keplerian potential).
Thus for elongated orbits
 eccentricity has decreased, which means also that the motion
is more circularly biased. The adiabatic analytical models already cited
predict indeed a moderate tangential anisotropy.
This trend for rounder orbits is also present in numerical results by
Norman \etal (1985), who studied how the shape of orbits close to 
a growing point mass is modified.

Initial anisotropies in the stellar velocities appear to have
a weaker influence than an initial cusp (Quinlan \etal
1995). Due, however, to the small number 
of available analytical anisotropic models,
 this problem could not be thoroughly examined.
One of the few anisotropic cases for which the cusp slope has been
 semi-analytically evaluated is a spherical model fully made of
 circular orbits. 
From conservation of energy and angular
 momentum, the final logarithmic slope is found to be
$\gamma=9/4$ (Quinlan \etal 1995
\footnote{
 using their notations in which
initially $\rho(r)\sim r^{-C}$,
the value for the slope given
in their Eq. (17) should actually be
 $C= 3 - \frac{3}{4-\gamma}$
}), significantly higher than for the isotropic
case ($\gamma = 3/2$).
Therefore we could expect that in a model with some intermediate degree
of tangentially biased motion, $3/2 < \gamma < 9/4$.  Since our
 axisymmetric models are flattened in their centre by an excess of
tangential motion, it is interesting to ask whether such an
anisotropy will have a sizeable influence on the cusp slope.
We may also ask whether a systematic rotation will bring
any changes. These are some of the questions we will try 
to answer in this paper. 

For the very few galaxies which had been observed ten years ago
with sufficient central resolution,
 observations  showed rotation velocities
 rising towards the center. This prompted attempts to compute the amount
of angular momentum brought to the center by the forming cusp (Lee \& Goodman
1989, hereafter LG89 ; see also Cipollina \& Bertin 1994,
 hereafter CB94). These
works, however, had to assume a spherical potential in order to 
derive analytical solutions.
One aim of this paper is to address this problem anew
 without any such approximation. In sections 5 \& 6, 
we investigate the influence of initial rotation
for self-consistent axisymmetric models.
Our motivations do not follow those of LG89, because higher quality,
more recent observations no longer show evidence for more
rotation than dispersion in the centre (see e.g. van der Marel \etal
 1997, Joseph {\it et al.} 2000 for HST observations of M32, and Kormendy \etal 1998,
regarding NGC 3377).

\subsection{Numerical methods}
\label{ss_num}

The pioneering simulation by Norman \etal (1985) did not
search for any detailed feature in the center, but 
 was meant instead to measure the shape changes
 caused at moderate and large radii by a central growing BH.
Due to the limited numerical capacity available at the time, their
results displayed  
substantial particle noise, but showed already the trend towards
axisymmetrization.
 Merritt \& Quinlan (1998) recompute this 
problem with a considerably improved accuracy, obtained by using
a larger number of particles ($N\sim 1.1 \times 10^5$ instead of 
$N= 2 \times 10^4$ for Norman \etal 1985), and
by enhancing the central particle density. To this end,
they replace each strongly bound particle by a number
of less massive particles distributed over its orbit
(see \S \ref{ss_fs}).

Sigurdsson \etal (1995) built a numerical code to follow the growth of the
 cusp itself, and tested it for the spherical case.
Their code makes use of the self-consistent field method
 (Hernquist \& Ostriker 1992). This 
 is formulated as a perturbation method, in the sense
that it uses a family of density-potential functions, of which
the zero-order term corresponds, or is close, to the initial model.
At each time step during the simulation, the coefficients
of the expansion for the density yield the new potential.
The less the system evolves, the fewer are the terms
required in the expansion to provide a given level of accuracy.
In the implementation by Sigurdsson \etal (1995),
 the number of particles in the central parts is enhanced 
by using a mass spectrum for the particles. 
 This is achieved by considering for the particles distribution, instead 
 of the initial d.f. $f_0(E)$, 
a distribution $f_0(E)/ m(E,L)$, such that the mass $m$ of a particle
is less than unity if the pericenter of its orbit is smaller than some
 radius, and is unity if it is larger.
This way, orbits reaching close to the centre are
represented by more numerous, lighter particles.
A numerical simulation based on this code, and using
$512,000$ particles, has been performed by
van der Marel \etal (1997), in order to check the stability 
of their dynamical model for M32, i.e. an axisymmetric model
including a central supermassive BH. 

In the present paper, we use a numerical perturbation 
technique which allows to concentrate particles at the center
of the model, without entailing at the same time high particle-noise
for the potential at larger radii (\S 4). In addition,
 we perform a subdivision of the central particles when
necessary, to further increase the resolution at the center.
 These numerical simulations are used here in order to
extend the models of BH induced cusps
 to non-spherical systems. 

\section{The initial axisymmetric model}
\label{sec:initial}

\subsection{Axisymmetric, non-rotating model}
\label{ss_f0}

Fully analytical axisymmetric models for 
elliptical galaxies are scarce (Lynden-Bell 1962, Dejonghe 1986, Evans
1994,
 {\it cf.} discussion in Hunter \& Qian 1993).
Most of them are sums --often infinite sums --
of Fricke terms.  We take here the simple model by
Lynden-Bell (1962), which uses only two Fricke terms:
 
\begin{eqnarray}
 f_0(E,L_z) =& F_1 E^{7/2} + F_2 L_z^2 E^{13/2} &~~ {\rm if} E>0\\
   ~  =& 0 ~~~~~~~~~~~~~~~~~~~~~~~ & ~~ {\rm otherwise} \nonumber
\label{eq_f0}
\end{eqnarray}

\noindent
where $E$ is the binding energy $E=\Phi_0 - v^2/2$ and $L_z$ is the
component of angular momentum about the symmetry axis.
This distribution corresponds to a flattened Plummer potential:

\begin{equation}
\Phi_0(R,z) = \frac{M_0}{((R^2 + z^2 + a_P^2)^2 - 2 b_P^2 R^2)^{1/4}},
\label{eq:pot0}
\end{equation}

\noindent
where we have used cylindrical coordinates.
In the above, $M_0$ is the total mass of the model, $a_P$ and $b_P$
are two parameters, and $F_1$ and $F_2$ are 
normalizing numerical constants, which expression is given by
Lynden-Bell (1962):
\begin{equation}
\begin{array}{l l}
 F_1 &=  \frac{\sqrt{2}} {4 \pi^{3/2}} \frac{5 !}{7/2 !} \frac{3 a_P^2 -
 2 b_P^2}{M^4} \nonumber \\
 F_2 &=  \frac{\sqrt{2} }{4 \pi^{3/2} }\frac{9 !}{13/2 !}
\frac{5 b_P^2(2 a_P^2 - b_P^2)}{M^8 }\\
\end{array}
\label{eq:F12}
\end{equation}
We take in the following $a_P=4$ and $M_0=1$ in the units of the
 simulation. These
are such that the mass unit is $10^{11} \,M_\odot$, the length unit
is $1\,$ kpc, and $G=1$.

The core radius is usually defined as the radius at which
the surface brightness $\Sigma(R,z)$ has decreased to half its central
value $\Sigma_0$.
This corresponds for the present model to $R_{1/2}\simeq 0.76 ~a_P$.
The mass enclosed within the $\Sigma=cst$ isodensity level passing at $(R_{1/2},z=0)$
is $M_{core} \simeq 0.2 ~M_0$.  

\begin{figure}
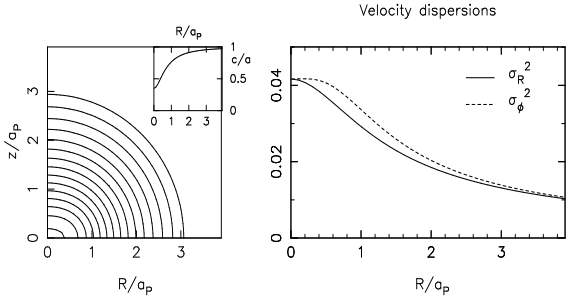

 \hbox{
 \includegraphics{figures/lyn_b1.ps}
 \hskip 3.8 cm 
 \includegraphics{figures/lyn_b2.ps}
 }
 \vspace{5cm}
 \caption{\label{fig:lynden} Left panel: isodensity contours of the Lynden-Bell model with $a_P=4$ and
 $b_P=2.2$; the axis ratio for the density is shown as a function of
the cylindrical radius in the insert.
Right panel: velocity dispersion profiles $\sigma_R^2 = \sigma_z^2$ 
and $\sigma_\phi^2$ along the $R$ axis for the
same model}
\end{figure}

The flattening decreases with radius, so that the models tend towards 
spherical symmetry at large radii. The larger $b_P$,
 the larger the central flattening. 
 $b_p$ however must be smaller than $ b_{P~max} \simeq 0.64 \, a_p$
 otherwise the isodensity contours have a dimple on the z-axis.
 We shall consider a model close to the model with 
 maximum flattening, by taking $b_P = 0.55\,  a_P $. For this model
the axis ratio is about constant, with value $c/a\simeq 0.55$
 (Fig. \ref{fig:lynden})
 within the region where the cusp will develop (see above
 the definition of the influence radius), 

 Since the d.f. depends only on $E$ and $L_z$, 
 the velocity dispersions obey $\sigma_R= \sigma_z$.
 The flattening is sustained by an excess of motion in 
the $\phi$ direction, with respect to motion in the $R$ or $z$ directions.
 If velocity fields that are moderately biased towards tangential
 motion can affect the final density cusp, we should witness it in our computations.

 The distribution function in Eq. \ref{eq_f0} is even in $L_z$ 
and corresponds to a
 non-rotating system. In this case ({\it cf.} Lynden-Bell 1962):
 \begin{eqnarray}
 \sigma_R ^2 =&  \frac{ \Phi_0}{6} \left( 1 - 0.2 \frac{C R^2 \Phi_0^9}
 {\rho_0}\right) \, = \, \sigma_z^2 \\
   \sigma_{\phi}^2 =& \frac{\phi_0}{6} \left(1 + 0.4 \frac{C R^2
 \Phi_0^9} {\rho_0}\right) & \hfil ,
  \label{eq:sigma}
 \end{eqnarray}
 with $C = 5 b_P^2 (2 a_P^2 -b_P^2)/(4\, \pi M_0^8)$.


Fig. \ref{fig:lynden} displays the isocontours for the density, as well
as the profile along $R$ for the  
velocity dispersions along each direction.

\subsection{Rotating models}
\label{ss_f0rot}

Models with positive net rotation are obtained 
by changing the sign of $L_z$ for a fraction of the orbits 
having negative $L_z$. If $\eta$ is the fraction of orbits 
being reversed, we obtain:

\begin{equation}
f^r(E,L_z) = \left\{ { f_0(E,L_z) (1-\eta)  ,\quad  L_z<0 }
    \atop { f_0(E,L_z) (1+\eta) , \quad L_z >0}  \right .
\label{eq:fr}
\end{equation} 
\noindent
In order however to avoid a discontinuity at $L_z=0$, the
distribution function must 
have $f(E,L_z)=f(E, -L_z)$ for $L_z \rightarrow 0$.
We therefore take for initial model:
\begin{equation}
 f_0^r(E,L_z) \equiv f_0(E,L_z) \times (1 + \eta p(L_z)),
\label{eq:defrot}
\end{equation}
 where $p(L_z)$ is some function having value $\pm 1$, 
except for a small range around the origin,
 where it goes to zero. A simple choice is:

\begin{equation}
p(L_z)= \left\{ \begin{array}{l r} 
{\rm sin}\left( \frac{\pi}{2} \frac{L_z}{L_m}\right) , & |L_z| < L_m \\
  {\rm sign}(L_z), & {\rm otherwise}  
\end{array} \right . 
\label{eq:plz}
\end{equation}
Here $L_m$ is a parameter which determines
 the extent of the $L_z$ region where rotation
is smaller than $\eta$. Its choice is of special
 interest for what follows. 

The simplest choice for $L_m$ is a constant, independent
of energy ({\it cf.} CB94).
However, by taking some function $L_m(E)$, we can build models
where, for any given energy, a constant fraction of orbits contribute
 to rotation. This is similar to the case studied by
LG89. We investigate both types of
models, as we know from comparing these two papers that
 they yield different results.

We thus consider two kinds of initial distributions, that use respectively:
\begin{eqnarray}
p_1(L_z) =& p(L_z)&\hskip -4pt {\rm with} \, L_m =  L_0  ~~~~~~~~~~~  \cr
p_2(L_z,E) =& \hskip -4pt p(L_z;E)& \hskip -4pt {\rm with}\, L_m = x L_c (E), ~  0 < x < 1 ~
\label{eq:defp}
\end{eqnarray}
In the above definition, $L_c(E)$ is the angular momentum for the
circular orbit with energy $E$, whereas $L_0$ is some constant,
 which is conveniently expressed in units of
$l_0= a_P \sqrt{\Phi_0(0)}$.

The mean tangential velocity is given by:
\begin{equation}
v_{\phi} (R,z)=\frac{ \int  d^3 v \, f_0^r(E,L_z) (L_z/R)}
                          {\int  d^3 v \, f_0^r(E,L_z) }. 
\label{eq:vphi}
\end{equation}

The effective rotation is measured by:
\begin{equation}
\eta_e= \frac {\int d^3r\, d^3v\, f(E,L_z) L_z} {\int d^3r\, d^3v\, f(E,L_Z)
|Lz|},
\label{eq:eta}
\end{equation}
which is in practice slightly smaller
than $\eta$  in the definition of $f_0^r$ given by Eq. (\ref{eq:defrot}) 
(and exactly $\eta$
in the limit $L_m \rightarrow 0$).

\section{Numerical method}
\label{ss_numrec}

A crucial feature for any $N$-body code aimed at studying the density cusp
is of course the central resolution. Most particles should ideally
orbit within a few times the `influence radius' $r_I$ of the BH
(Eq. \ref{eq:ri}). In this paper, we use a scheme which is formulated
 as a perturbation
method, and allows to choose freely the sampling in phase-space.
The main advantage of our scheme is that even if we have few particles
outside $r_I$, where evolution is weak, the potential
is fairly well represented (since it is 
dominated by the analytical $\Phi_0$ term, which has zero particle-noise).

\subsection{The `perturbation particles' scheme}
\label{ss_PP}

The PP scheme has been described in detail elsewhere
 (Leeuwin {\it et al.} 1993). In this scheme, one writes
the collisionless Boltzmann equation (CBE) as a perturbation problem.
The unperturbed state corresponds to the initial analytical model,
 while the perturbation corresponds to
the evolution of the system.
Thus, the system d.f. is formally written as
$f= f_0 + f_1(t)$, where
$f_1(t)$ is found by integrating explicitly
the CBE along orbits in an N-body realization.
The choice of the orbits to be integrated during the run --that is to
say the initial distribution of particles $f_S({\bf r},{\bf v})$--
is in principle arbitrary, but one should aim at a good sampling
of phase-space regions where
 large perturbations are expected.
The function $f_S$ is chosen to be a function of the
system classical integrals (here $E,L_z$), so that in
the absence of time evolution the sampling of
phase-space is stationary.

If the system potential varies with time, $f_S$ can not be made
 a stationary function. However, 
 both $f_S$ and $f$ obey the CBE equation.
 Therefore, noting the phase-space coordinate
$w\equiv ({\bf r},{\bf v})$, we may simply record
the initial value $f_S(w(t_0))$ for each particle, since along each orbit
$f_S(t)=f_S(t_0)$. 
The mass attributed to the particle running on that orbit
is weighted according to the local phase-space density:
\begin{equation}
m_1(t)=\frac{ f_1(w(t)) }{f_S(w(t_0))},
\label{eq:m1}
\end{equation}
as explained in Leeuwin {\it et al.} (1993) and Wachlin {\it et al.} (1993).

On the other hand, the CBE is explicitly solved for $f_1$: 
the accuracy of orbits integration can then be monitored through
the relative variations of $f$.
The perturbation along an orbit is given by:

\begin{equation}
\frac{{\rm d} f_1(t)}{{\rm d}t}=- \left[ f_0,
  \Phi_{BH} + \Phi_{s.g.}\right] 
\label{eq:df1}
\end{equation}

\noindent
where $\Phi_{BH} = M_{BH}/r$, and $\Phi_{s.g.}$ is the self-
gravitating potential of the cusp.

The computation of moments for $f_1$ amounts in fact to a 
Monte-Carlo integral with sampling $f_S$. 
For instance, the perturbed total mass is evaluated as:
\begin{equation}
 M_1 \equiv \int f_1(w)~ d^6w \simeq
 \Sigma_{i=1}^N \frac{f_{1}(w_i)}{f_S(w_i)} \simeq \Sigma_{i=1}^N m_{1i}. 
\label{eq:mass1}
\end{equation}
To compute local values of the density field, or of any other field,
the sum over the particles is of course limited to a small
 region of space where the quantity of interest is
roughly constant. Moments of the total distribution $f=f_0+f_1$ are given by:

\begin{equation}
 <V^n> = \frac{ \int (f_0+ f_1)~V^n ~d^6w}
{\int (f_0 +f_1)~ d^6w},
\label{eq:moments}
\end{equation}

\noindent
where $V$ represents any component of the velocity vector.
This is estimated as:

\begin{eqnarray}
<V^n> ({\bf r}) &= &\frac{ \int f_0~ V^n ~d^6w ~+~ \int f_1~ V^n ~d^6w}
{\rho_0 ({\bf r}) + \rho_1({\bf r})} \nonumber \\
 &\simeq & \frac{ \rho_0 ~<V^n>_0 ~ +~ \Sigma_i
\frac{f_{1}(w_i)}{f_S(w_i)} V^n_i }{\rho_0 + \rho_1}
\nonumber\\
& =&\frac{ \rho_0
~<V^n>_0 ~ +~ \Sigma_i m_{1i} V^n_i }{\rho_0 + \rho_1}
\label{eq:vel}
\end{eqnarray}

The sampling distribution $f_S$ is most efficient for the evaluation of the
above Monte-Carlo integrals, when it is most similar to the integrants.
The integration accuracy can be checked by monitoring the total mass
 in the perturbation, which should remain equal to zero. A
convenient measure for the associated error is
 $M_1/||M_1||$, where we note 
$||M_{1}|| \equiv \Sigma_i |m_{1i}|$.

The self-gravity of the cusp corresponds to the force
due to the perturbation mass. Thus, a particle $j$ will be subject to
a self-consistent force term:

\begin{equation}
\nabla \Phi_{s.g.} ({\bf r}_j) \simeq \Sigma_{i\ne j} \frac{m_{1i}}{|{\bf r}_i-{\bf r}_j|^3}
 ({\bf r}_i-{\bf r}_j).
\label{eq:selfg}
\end{equation}

\subsection{Orbit integration}
\label{ss_orbits}

Close to the BH, the force gradients are of course very large. Some
orbits therefore demand extremely short time-steps during small
time intervals.  For such problems, adaptative time-steps is the most
convenient scheme. 
In a 1D simulation (Leeuwin \& Dejonghe 1998) that 
we ran to check whether the PP method could be advantageous for the
present problem, we have experimented block time-steps integration 
(using either a leap-frog, or a RK4 scheme).
In a block time-step scheme, the particles are sorted into groups at regular
time intervals, and each group advanced with its
smallest time-step until the next sorting out. 
This makes the scheme a costly one,
since a set of particles may be advanced with a needlessly tiny timestep.

Symplectic methods (Wisdom \& Holman 1991, Saha \& Tremaine 1992),
because they conserve phase-space volume,
would in principle be useful to guarantee conservation of our sampling
phase-space density (Binney, {\it private communication}).
Unfortunately, symplectic integrators 
with adaptive time-steps do not yet exist (see Duncan, Levison \& Lee
1998 for a version with block time-steps).

We use in this work the {\small ODEINT} routine due to 
Press {\it et al.} (1986), which is an adaptative individual
time-steps scheme, based on a 4th-order Runge-Kutta time
interpolation. The conservation of $f$ along orbits is better
than $10^{-4}$ over each entire simulation.

\subsection{Force computation with \grape  machines}
\label{ss_grape}

The BH mass was updated at each time-step.
This is straightforward since the time dependence is explicit.
Moreover, within the BH influence radius where
 the BH force term dominates, a precise evaluation is required.

The self-gravitating force (Eq. \ref{eq:selfg}) is computed
by direct summation of the individual particles contributions, using a \grape
machine. A small modification of the standard software was necessary since the
PP masses can be either positive or negative, a possibility not
foreseen for \grape.
Furthermore, timesteps within the cusp
are very small, so that the cost of re-evaluating by direct summation
the inter-particles force at each time-step for the most bound orbits
would be prohibitive. Since the BH grows very
slowly ($\sim 10^7$ Yrs) compared to the central dynamical times ($\sim
10^5$ Yrs, see \S \ref{ss_analytic}),
the force field due to the cusp also changes slowly 
with respect to individual crossing times.
Moreover, it is
negligible at large radii where dynamical times are larger
 (see Fig. \ref{fig:dm1}). 
Therefore the self-consistent force field was evaluated using \grape only
at regular time intervals $\Delta t$. To derive the self-gravity
force experienced by each particle at intermediate times, we
proceed as follows:
\begin{itemize}

\item The self-consistent forces are evaluated
using ~\grape~ 
at regular time intervals $\Delta t$. 
This yields the three 
components $(F_{xi}, F_{yi}, F_{zi})$ of the force 
for each particle $i$. 
We infer for each particle the two components 
$(F_{Ri}, F_{zi})$ along $R$ and $z$ in cylindrical coordinates.

\item Using a cell-in-cloud (CIC, {\it cf.}
Hockney \& Eastwood 1981) scheme in polar coordinates,
 we compute the force field $(F_R,F_z)$ on a grid in $R,z$
(we average over the azimuthal angle $\phi$).  

\item At each time-step within $ \Delta t$,
 the force $(F_R,F_z) (r_i(t))$ at
 the current position of each particle
 is interpolated from the grid values (with again a CIC).

\end{itemize}

The grid has $20 \times 20$ cells, and 
 extends in both directions from $x_1= r_I /200$ to $x_{20}=4\, a_p$.
Points are spaced with a logarithmic increment: 
\begin{equation}
x_{i+1}= x_i + dx_{i}\, k^a,
\label{eq:gridr}
\end{equation}
where $a$ and $k$ are two real parameters numerically determined after the
 choice of $x_1, dx_1$ and $x_{20}$. We have taken $dx_1=x_1$. 

For the force evaluation with \grape we use a softening length
$\epsilon= 10^{-4} \, a_P$.
 The force is computed $1000$ times
 during the BH growth. At each time-step,
the value for the self-gravity force is linearly interpolated from
the grid values, using again a CIC scheme.

A run using $\simeq 3\times 10^5$ particles, and the
parameters for the standard case defined hereafter (\S 5.1), takes 
approximately $2$ days on the Marseille 5 board GRAPE-3AF system, 
and somewhat longer on the GRAPE-4 system with 41 chips (see
Athanassoula
 {\it et al.} 1998 for the characteristics of the Marseille \grape
-3AF system,
and Makino {\it et al.} 1997, Kawai {\it et al.} 1997 for the
characteristics of the \grape-4 system).
For $N=300,000$ particles, one force evaluation by \grape-3AF takes 
approximately $140$ s, therefore $\sim 38$h in the simulation are
devoted to the computation of the self-gravity force field.
 
\subsection{Choice of the sampling distribution}
\label{ss_fs}

In order to ensure a stationary distribution (in the absence of any
perturbation), $f_s$ must be a function of the isolating integrals of motion
for the unperturbed potential $\Phi_0$.
We are limited in the choice of analytical distribution functions $f_s(E,L_z)$,
as we were for $f_0(E,L_z)$, because such models are scarce in the iterature.
The known d.f.'s for axisymmetric systems are in general expressible as
Fricke series, for which the self-consistent density can be calculated.
The simplest choice is thus to consider a term from a Fricke series

\begin{equation}
f_s(E,L_z) = F\, E^\alpha\, L_z^\beta,
\label{eq:fsF}
\end{equation}
\noindent
 where $E=\Phi_0 - \frac{v^2}{2}$ and $F$ is a normalizing constant.
 The corresponding density (see {\it e.g.} Dejonghe \& de
Zeeuw 1988) is given by:

\begin{equation}
\rho_s(R,z)= A\,F\, R^{\beta}\, \Phi_0^{\alpha +\beta/2 + 3/2},
\label{eq:rhos}
\end{equation}
 with $A=\frac{\Gamma(\alpha +1)\Gamma(\beta/2+1/2)}{\Gamma(\alpha+\beta/2
+1/2)}$.
We can not generate a density of particles with a strong cusp
using such a term. Models with a cusp in $R$ can in principle be built by
choosing $-1< \beta <0$, however (i) their cusp
in $R$ has a rather faint logarithmic slope $-1 <\beta<0$;
(ii) these are d.f.'s with
a cusp in $L_z$, so their sampling of velocity space is very
 inhomogeneous. This is actually less unfavourable than it seems, since
it favours nearly-radial orbits, which are actually those which
 experience the strongest perturbation (see \S \ref{ss_dfBH}).
More freedom is available regarding the density decay at large radii.
A steeper decline of density, with respect to the unperturbed model,
follows from taking $\alpha > 7/2 + \beta/2 $, as can be easily verified.
We are not worried by the low corresponding particle density
 at large radii, since 
the potential there will be dominated by the analytical term.

\begin{figure}
\vskip 6cm
\includegraphics{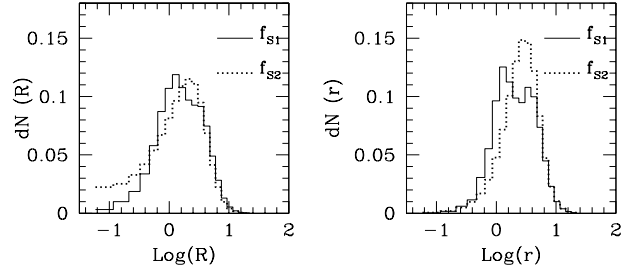}
\vspace{5cm}
\vskip -7cm
\caption{\label{fig:pnum}
Histogram for the normalized distribution of particles for each of
the two sampling distributions $f_{S1}$ and $f_{S2}$ discussed in the
text, as a function of the cylindrical radius $R$ (left) and of the
spherical radius $r$ (right)}
\end{figure}

Here we experiment with two very different Fricke terms for $f_s$,
 in order to make sure that the choice of the sampling does not
affect the measure of the cusp slope, which is our main objective. We use:
\begin{enumerate}
\item $f_{S1}$: $\beta=0,\, \alpha=8$.\\
This is a model with a core, but its density falls much faster than
the mass density of the
unperturbed model.
The central particle density is enhanced by splitting particles
according to their energy and angular momentum,
in a way similar 
to Merritt \& Quinlan (1998). First particles
are binned according to their binding energy $E_0$. For the most bound
particles, we evaluate their pericenter $R_{min}(E,L_z)$. If
$R_{min}$ is smaller than the final BH influence radius (see \S 4.3),
the orbit is selected for splitting. It is then
integrated, within the unperturbed potential,
 for sufficient dynamical times. A number $n$ of phase-space positions
along the orbit are recorded. Then the initial particle $(r(t_0),v(t_0))$
 is replaced by $n$ particles, each of which is given one of the
$n$ recorded positions and a statistical weight $\Gamma= 1/ 
[n\, f_S(r(t_0),v(t_0))]$. Usually, out of
20 equally sized bins in energy, we consider for splitting
 the $j=17 \dots 20$ bins,
take  $n=(j-17)^2$, and repeat the whole procedure a second time
(with $n=(j-15)$).
\item $f_{S2}$: $\beta =-0.9,\, \alpha= 7$.\\
This choice ensures a central density with cusp $\propto R^{-0.9}$,
which proves an effective way to increase the central number of
particles. We do not perform any splitting.
On the other hand the sampling is very inhomogeneous in
$R$ and $z$, as already mentioned.
\end{enumerate}
The two particles distributions arising from the two sampling distributions
just described are displayed on Fig. \ref{fig:pnum}.

\section{Cusps in non-rotating models}
\label{sec:results}

\subsection{Cases considered}
\label{ss_cases}

We assume a point mass is growing at the center of the model,
by gas accretion, without depleting the stellar component
(as in Young 1980, Quinlan \etal 1995, CB94, LG89).
The BH mass grows over a time $t_{BH}$ from $M=0$ at $t=0$,
 to the final value $M_{BH}$ 
following ({\it cf.} Merritt \& Quinlan 1998):
\begin{equation}
M (t) = M_{BH} \,\left( \frac{t}{t_{BH}} \right) ^2 \left( 3 - 2
\,\frac{t} {t_{BH}} \right).
\label{eq:mbh}
\end{equation}
\noindent
For most of the runs discussed hereafter, we took $t_{BH}=10^7$ Yrs.

The potential due to the BH is modeled using a Plummer potential with
smoothing length $\epsilon_{BH}$:
\begin{equation}
 \Phi_{BH}= M/\sqrt{r^2 + \epsilon_{BH}^2}.
\label{eq:potbh}
\end{equation}
We can choose $\epsilon_{BH}$ by imposing e.g.:
$\epsilon_{BH} \ll r_I$. For instance, $\epsilon_{BH} \leq 10^{-3}
r_I$. For the standard case, where $r_I \simeq 8 \times 10^{-2}$, we have
taken $\epsilon_{BH} = 5\times 10^{-5}$, and verified that this value was small
enough (\S \ref{ss_check}).

We consider as the standard case one where the BH has a mass
of $2\%$ the galaxy mass. This is supposed to be roughly representative 
of observational data. However, our flattened Plummer model
does not have a realistic density profile, so that a more
meaningful figure may be the ratio of the BH mass to
the initial core mass. The latter is roughly $0.2 \, M_0$ (see
\S \ref{ss_f0}). 

In addition, we discuss below experiments with higher BH masses, 
and different growth times. The summary of all the cases considered is
to be found in Table \ref{tab:list_runs}.
We will also consider, in the next section, models having initially
 some rotation. Those will be summarized in Table \ref{tab:listrot} of that section.

\begin{figure}
\includegraphics{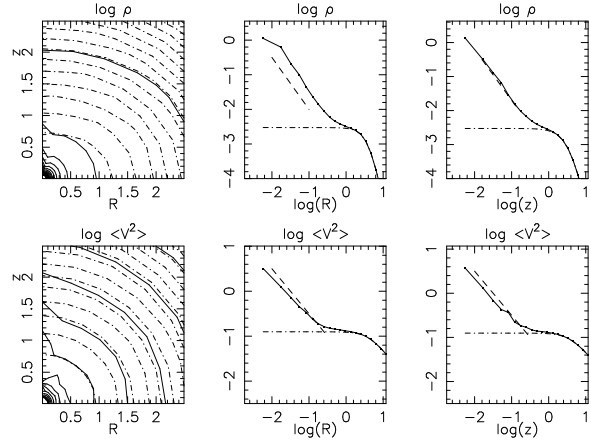}
\vspace{7cm}
\caption{\label{fig:ck0}
 The two left panels show iso- contours for the density
 (top) and the total velocity dispersion (bottom),
 for the final model in the standard run
 (solid lines). The levels are logarithmically
spaced. The initial model is shown by the dash-dotted lines (using
different logarithmically spaced levels).
The initial (dots) and final (full line)
 profiles along {\it resp.} the $R$ axis, and the $z$ axis,
are also displayed, using logarithmic scales. 
}
\end{figure}

We display on Fig. \ref{fig:ck0} the results for the standard case.
The sampling distribution used was $f_{S1}$.
For that case, and other ones where the BH mass is very small, 
statistical noise is reduced by adding $10$ snapshots taken within
a small time interval after $t_{BH}$. This is much less useful
for the runs with larger BH masses (see below, Table \ref{tab:list_runs}). 
A polar grid has been used to produce the graphs, with 
$20$ points logarithmically spaced along $r$ (based on
Eq. \ref{eq:gridr}), and $10$ points
equally spaced within $[0, \pi/2]$ along $\theta$, which is the angle from the
$z$ axis. 
The smallest value $r_{1}$ of the $r$- grid 
is chosen so that the total snapshot has at least $1,000$ particles
with $r<r_1$. 

Quantities associated to the particles correspond to perturbed quantities.  
The related grid quantities ${\bf a}_{ 1,kl}$
at grid points $(r_k,\theta_l)$
 are derived  by applying a CIC scheme in the polar geometry.
The total fields are finally recovered by adding the corresponding
 unperturbed analytical terms evaluated on the grid points. 
Thus for the total density:

\begin{equation}
\rho_{kl} = \rho_0 (r_k,\theta_l) + \rho_{1,kl}.
\label{eq:grid}
\end{equation} 

Similarly the velocity components and dispersions are derived
by adding the analytical terms, using Eq. \ref{eq:vel}, for
 $V=V_R, V_\phi,$ and $V_z$, and $n=1,2$.
Fig.  \ref{fig:ck0} shows isolevel contours for the total density at the
end of the simulation, as well as for the total velocity dispersion
$\sigma^2= \sigma_R^2 + \sigma_z^2 + \sigma_\phi^2$.
 The dash-dotted curves show the initial, unperturbed quantities. 
Also displayed (first row) are the curves $\rho_{k,1}$ and 
$\rho_{k,10}$, which give approximately the profiles of the total
density along $R$ and $z$, respectively.
 The same is drawn on the bottom row for the
square total dispersion $\sigma^2$. The dotted segments on these
graphs indicate a line with logarithmic slope
 respectively $3/2$ for the density, and $1$ for the square total dispersion.

\begin{figure}
\includegraphics{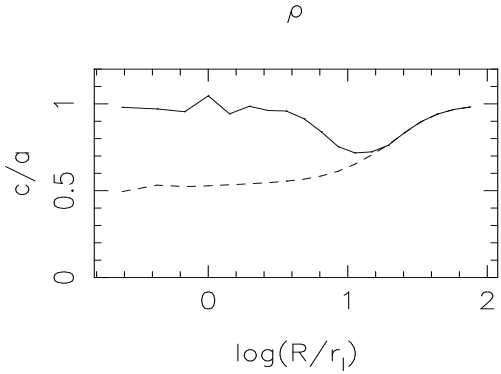}
\vspace{5.7cm}
\caption{\label{fig:ratio}
Minor to major axis ratio, as a function of $R$ 
scaled to the influence radius $r_I$, for the standard
run (solid line). 
The dashed line shows the axis ratio for the initial model.}
\end{figure}

Therefore, the figure shows that even for the less massive
BHs considered in this work, the central cusp is well resolved
in two-dimensions. The maps also show that the central region has become
very nearly spherical; the final axis-ratio is displayed on
Fig. \ref{fig:ratio} as
a function of the major-axis radius, scaled to $r_I$.
By eye inspection of \ref{fig:ck0}, the slope for the velocity is very similar 
to what is expected theoretically; also the density cusp appears
to be similar to what is predicted for an initially spherical model.
The cusp slopes will be measured with more care and discussed further 
in \S \ref{ss_slope}.
We now turn to some runs performed in order to check our computations. 

\begin{table}
\vskip 5mm
\caption{ Parameters for the `standard' run}
\begin{minipage}[c]{10cm}
\vbox{
\begin{tabular}{|l c|} 
\noalign{\hrule}
   &    \\
~Final $M_{BH}/M_0$ \dotfill & $0.02$ \\
~Time for BH growth  \dotfill & $10^7$ Yrs \\
~Smoothing length for the BH \dots &~~~ $5\times 10^{-5} a_p$ ~~\\
~Number of particles \dotfill & $\sim 300,000$\\
 & \\
\noalign{\hrule}
\end{tabular}
}
\end{minipage}
\label{tab:stand}
\end{table}



\subsection{Numerical checks}
\label{ss_check}

A run was first made for $b_P=0$, in order to check that
the slope obtained for a spherical potential agrees with
the analytical estimates for this case. 
Its slope is measured, in the way we will explain in \S \ref{ss_slope},
both for the density and the velocity dispersion. We measure respectively
$\gamma_{\rho} \simeq 1.52  \pm 0.02 $ for the density,
in good agreement with the adiabatic model (and the simulation by
Sigurdsson \etal 1995), and $\gamma_{\sigma^2} \simeq 0.94 \pm 0.05$,
very close to unity, as 
expected in the nearly Keplerian potential produced around the BH.

We also made some numerical checks for the standard
 axisymmetric run. First of all, the simulation
 was pursued after the BH growth time for an equal duration, 
in order to verify that a stationary model had been reached.
The influence of the smoothing length $\epsilon_{BH}$ 
used in the BH potential was tested, by comparing runs using either
a larger ($\epsilon_{BH}= 10^{-4} \, a_P$) or a smaller ($\epsilon_{BH}=
2\times 10^{-5} \, a_P$) smoothing length.
The larger value produced a slightly shallower cusp, but
results were unaffected for the smaller value.
Therefore we use $\epsilon_{BH}=5\times 10^{-5}\, a_P$ in all subsequent
runs --including those with larger BH masses, where a larger value of
 $\epsilon_{BH}$ could have sufficed.

\begin{figure}
\includegraphics{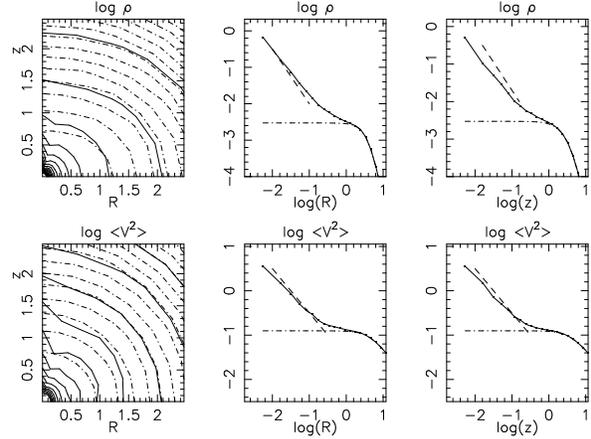}
\vspace{7 cm}
\caption{\label{fig:fs2}
Same as the previous figure,
also for the parameters of the standard case,
 but using $f_{S2}$ for the particles initialization, instead
of $f_{S1}$. The layout is as for Fig. \ref{fig:ck0}. }
\end{figure}

We have experimented with both the $f_S$ described in \S \ref{ss_fs}.
The final model obtained, for the parameters of the standard case,
but using $f_{S2}$ instead of $f_{S1}$, is shown on
Fig. \ref{fig:fs2}.
Results can be seen, by comparing to Fig. \ref{fig:ck0},
 to be extremely similar to those obtained when
using the other sampling distribution, in spite of the fact
that $f_{S1}$ and $f_{S2}$ are very different functions.
The first sampling ($f_{S1}$) turned in practice 
 to yield results with apparently somewhat 
less particle-noise for the density. 
 This is probably due to the fact that $f_{S1}$ corresponds to a 
particle distribution which is similar along the
$R$ and $z$ axes, while $f_{S2}$ has a cusp only in $R$. 
This may for instance explain why
the spherically averaged cusp has more mass for
$f_{S1}$ than for $f_{S2}$ (see Fig. \ref{fig:1d_ck0}). Some perturbation
mass may be less well sampled in the $z$ direction.
The differences, however, remain marginal. The two sampling functions give
 estimates for the logarithmic slope of the cusp (see \ref{ss_slope})
that can not be distinguished, within the error bars. 
Therefore we are confident that
our results are not affected by the initial particle distribution.

A few runs were also performed in order to make sure that the results do
not depend significantly on numerical parameters, such as the spacing
of the grid used for the force evaluation (\S \ref{ss_grape}),
 the total number of force computations by \grape (\S \ref{ss_grape}),
 or the choice that we made for the number of particles $N=3 \times 10^5$.

\subsection{Importance of the cusp self-gravity}
\label{ss_selfg}

The perturbed central densities are high, but within small
volumes. Therefore the mass in the cusp is not necessarily
very large. Applying Eq. \ref{eq:ri}
to the Lynden-Bell model we find for the influence radius:
\begin{equation}
r_I \simeq 2\, a_P \frac{M_{BH}}{M_0}.
\label{eq:riLB}
\end{equation} 
For the standard run $M_{BH}/M_0= 0.02$, thence $r_I \simeq 0.16$.
Initially, the mass enclosed in this radius is roughly (neglecting the
effects of flattening):
$M_0(r_I) \simeq  M_0 (r_I/\sqrt{r_I^2 +a_P^2})^3 \simeq 2\times 10^{-5}
M_0$. At the end of the
simulation, we find it is: $M(r_I) = M_0(r_I) + M_1(r_I) = M_0(r_I)
+ \Sigma_{(r_i<r_I)} \, m_{1i} \simeq 2 \times 10^{-5} + 6\times
10^{-4} \simeq 6\times 10^{-4} \, M_0$.
Therefore the mass within the cusp remains a small fraction of the
system mass.

\begin{figure}
\vskip 5.5cm
\hbox{
\includegraphics{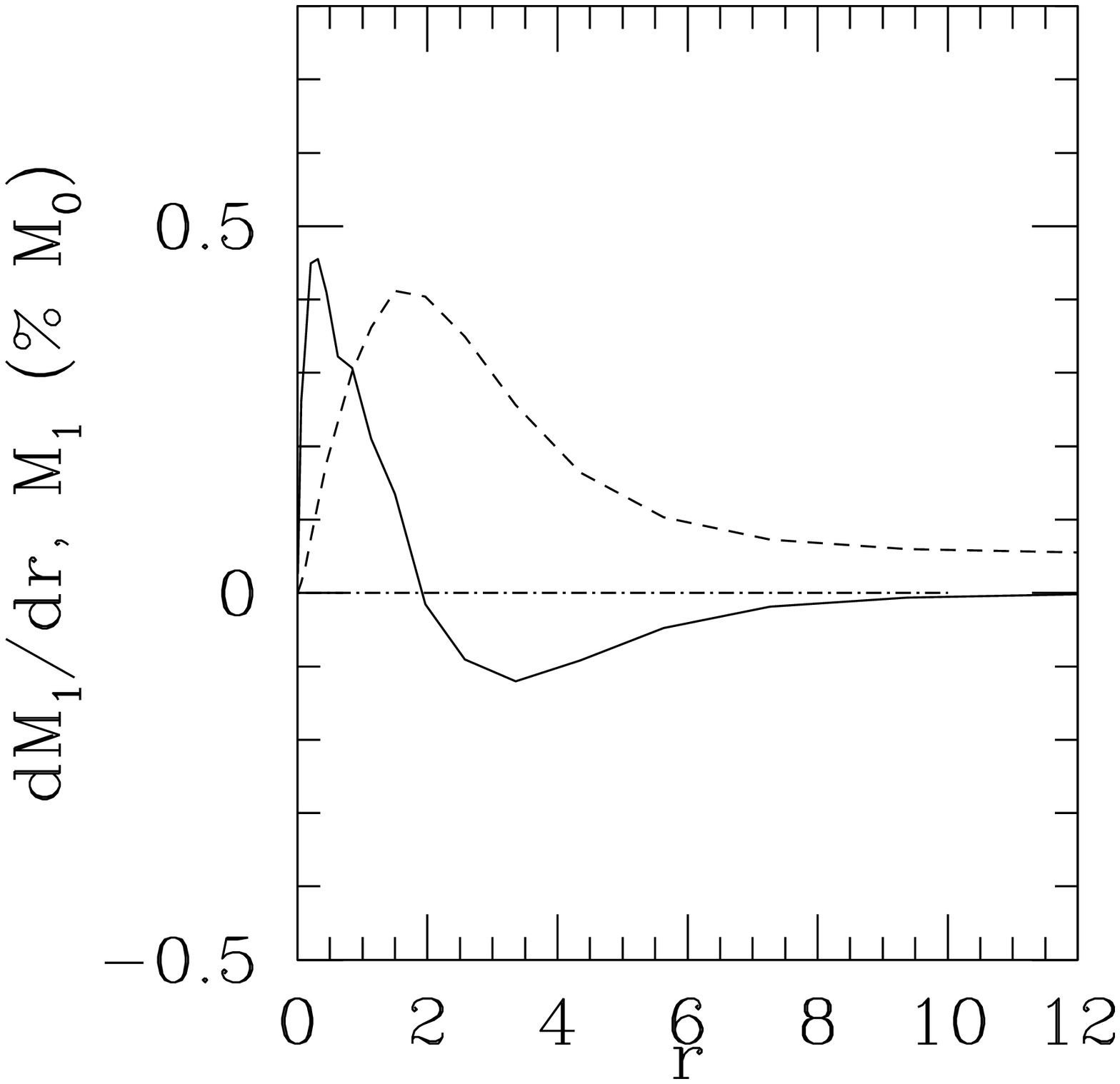}
\hskip 4.3cm
\includegraphics{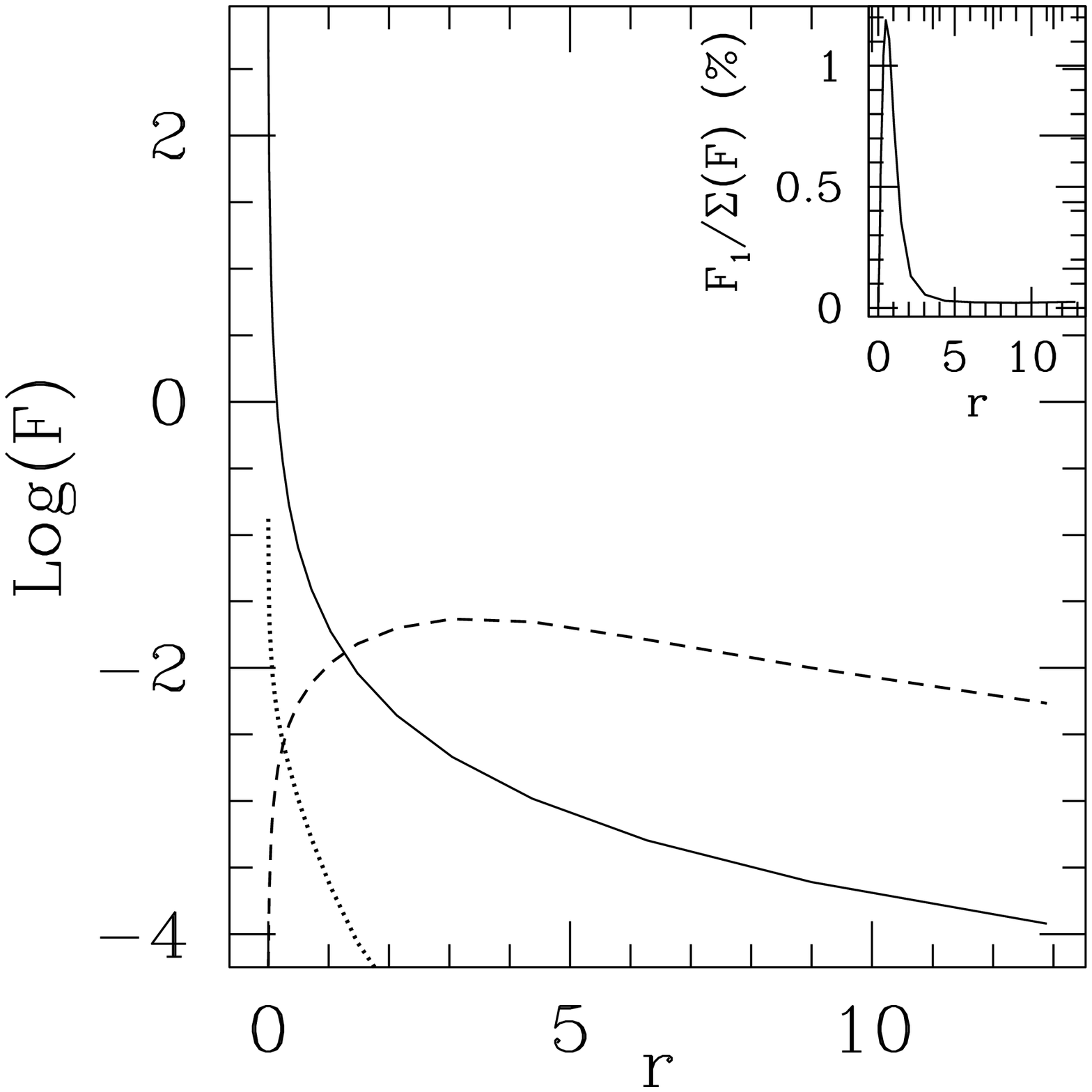}
}
\vspace{4.5cm}
\vskip -5cm
\caption{\label{fig:dm1}
Left: The radial distribution of perturbation mass $dM_1/dr$ 
(solid line) and the cumulated distribution $M_1(r)$ (dashed line),
for the standard case with $M_{BH}/M_0 = 0.02$, in percents of $M_0$. 
Right: Comparison between the forces due to the unperturbed
 mass distribution (dashed line), the BH (solid line), and the
cusp gravity (dotted line). The insert shows the ratio of the
cusp force $F_1$ to the total force $F_0+F_1+F_{BH}$, in percents.}
\end{figure}

On the other hand, the system has been affected by the central mass
in a region much larger than the cusp itself. This
can be seen from the radial distribution of the perturbation mass
density, displayed on Fig. \ref{fig:dm1} for the standard case.
For $r\geq 2$,  $\frac{dM_1(r)}{dr} <0$, 
corresponding to orbits that have been requisitioned in order
 to build the cusp;
this negative perturbation extends far from the center.
 As a consequence, the cumulated mass $M_1(r)$ decreases after
$r\simeq 2$. It is $10\%$ of its maximum value at $r\simeq 5$.
 Outside this radius the cusp force is practically zero (Fig. \ref{fig:dm1}).
The orbital
participation will be further described below (\S 5.5).

We may estimate the self-consistent 
force (due to the cusp itself) $F_{s.g.}=\nabla \Phi_1$ within the
influence radius, where we assume $\rho_1(R,z) \approx K \, r^{-3/2}$.
A reasonable rough guess for $K$ is such that
$\rho_1(r_I) \approx \rho_0(r_I)$. This yields a roughly correct
`calibration' for the central value, in agreement with the numerical value
at the end of the simulation. 
Fig. \ref{fig:dm1} displays the different radial force terms: 
$F_0=\nabla_R \Phi_0$, $F_1=\nabla_R \Phi_1$, $F_{BH}$.
Within a substantial fraction of the initial
core radius ($\sim 0.75 ~a_P=3$), the unperturbed force is negligible.
The BH dominates the central region, within
roughly $a_P/4$, by a factor scaling from
$10^4$ at $R\simeq 0$ to $10^2$  at $R \simeq r_I$.
The self-consistent force contribution reaches at most
$\sim 1\, \%$ for $M_{BH}/M_0 = 0.02$. Therefore we have
neglected it for most runs with this BH mass. 
 The self-gravity contributes
at most for $\simeq 2\%$ when $M_{BH}/M_0 = 0.03$ 
(see Table \ref{tab:list_runs}). This is still very small,
therefore we have also neglected the cusp self-gravity
for models with $M_{BH}/M_0=0.03$.

\begin{figure}
\vskip 5cm
\hbox{
\includegraphics{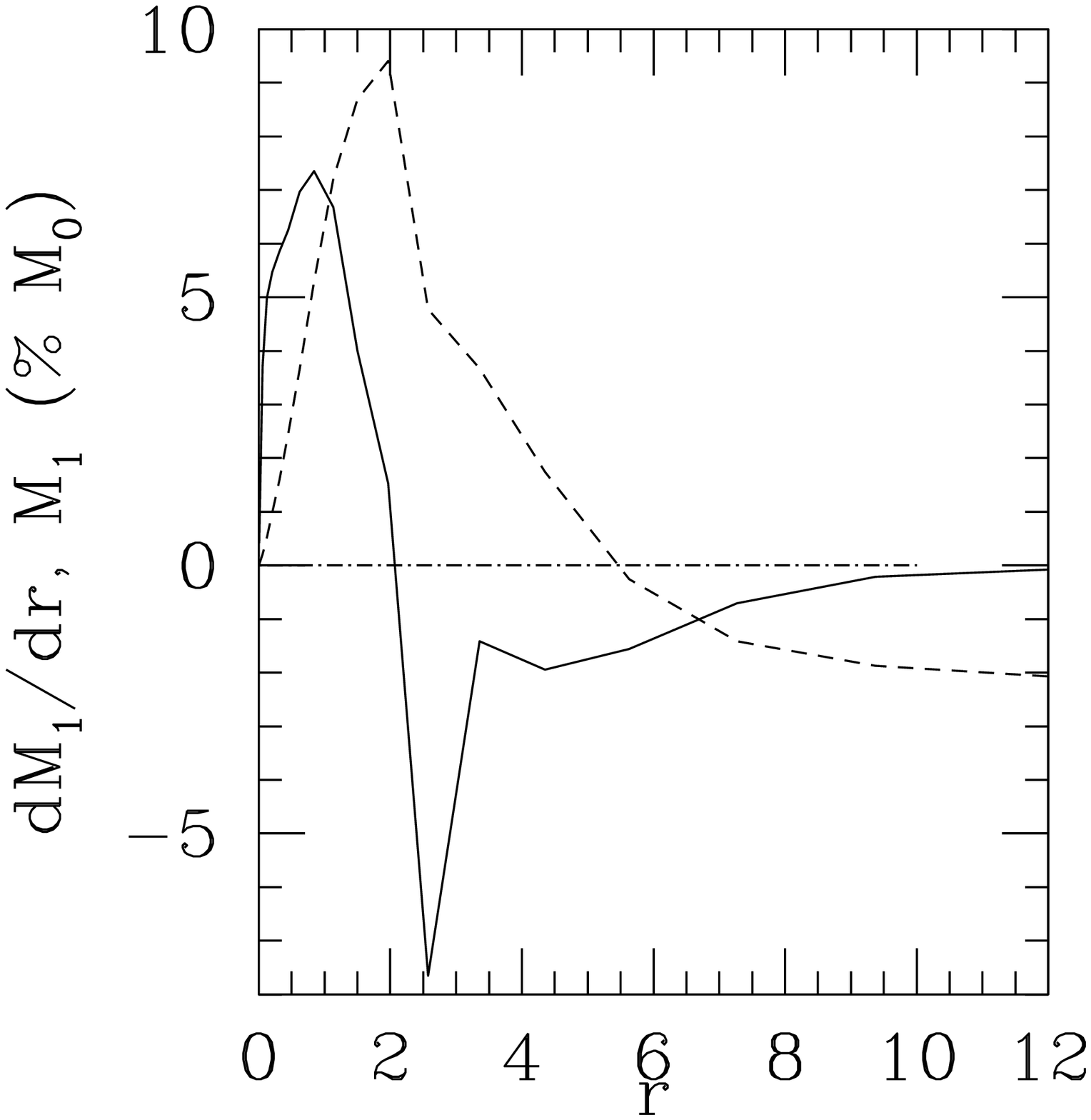}
\hskip 4.2cm
\includegraphics{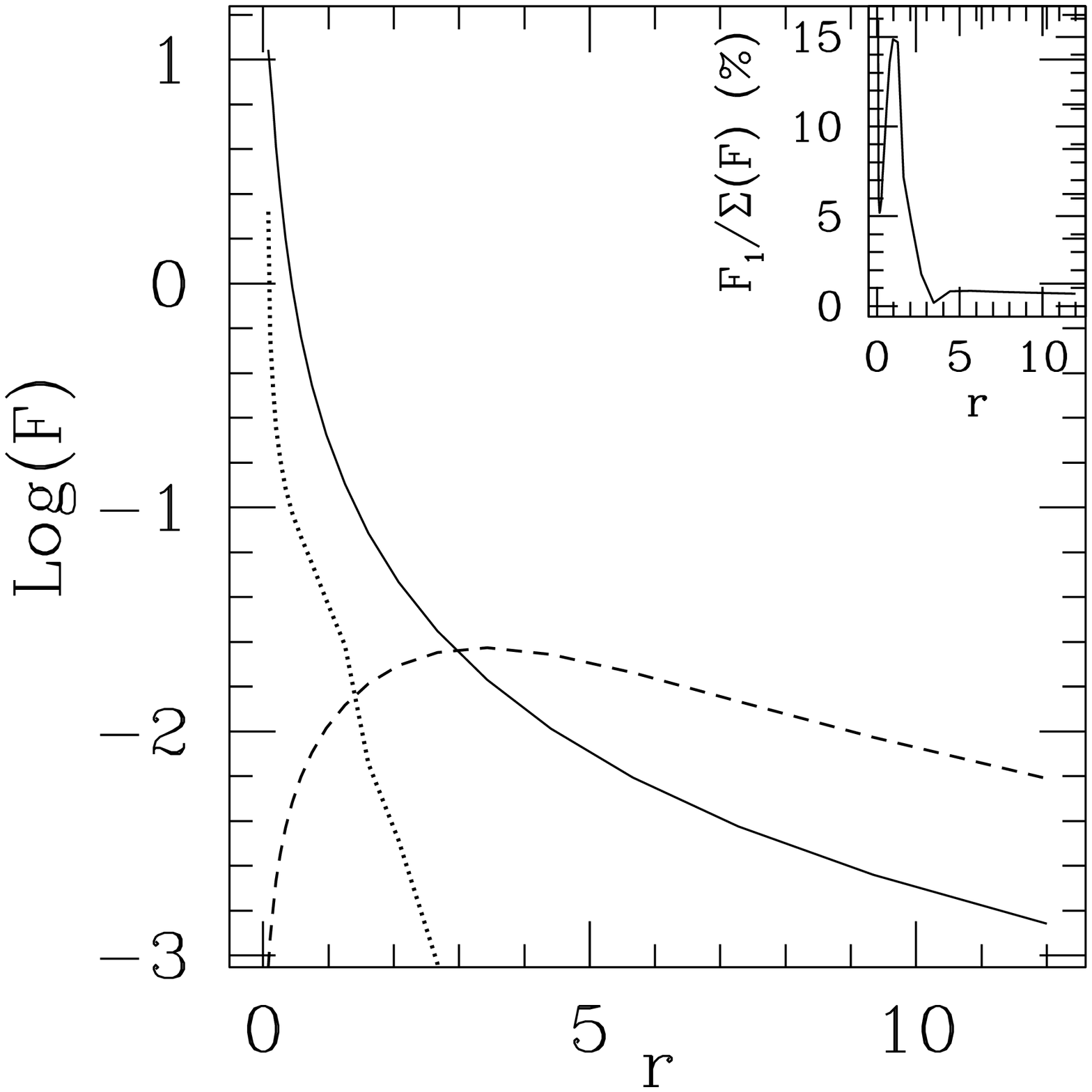}
}
\vspace{5cm}
\vskip -6cm
\caption{\label{fig:dm2}
Same as Fig. \ref{fig:dm1} but for
a BH with mass $M_{BH}= 0.2 \, M_{0}$}
\end{figure}

The self-consistent term is not negligible, however, for the
larger mass ratio $M_{BH}/M_0 = 0.2$ considered in this work,
since it contributes for up to $\sim 15\%$ of
the total force at $R \simeq a_P/4$ 
(Fig. \ref{fig:dm2}).

\subsection{Measure of the cusp slope}
\label{ss_slope}

The Poisson noise due to the discretisation into particles 
varies with the number of particles as $1/\sqrt{N}$. To
reduce it, we continue the simulation for another $0.5 \times 10^7$ Yrs, 
and draw 10 snapshots, at regular time intervals.
The snapshots are then merged, and this snapshot of $10 \, N \simeq
3\times 10^6$ particles is analyzed. This is especially useful for the 
runs with $M_{BH}/M_0 = 0.02 $, and, to a lesser extent,
$M_{BH}/M_0=0.03$.

The central region (within $\sim ~r_I$) is very nearly spherical,
 as can be seen from the 2D contour map of $\rho(R,z)$ (Fig. \ref{fig:ck0}).
Therefore, in order to measure the cusp slope, we can
integrate over the $\theta$ angle of the polar coordinates $(r,\theta)$. 
A logarithmic grid in $r$ is built, using Eq. \ref{eq:gridr} and 
such that the sphere with radius the first grid point $r_1$ 
encloses $\simeq 1000$ particles. The spherically symmetrized density
and velocity dispersions are evaluated at the grid points, and these grid
values used to derive the corresponding logarithmic slopes.

\begin{figure}
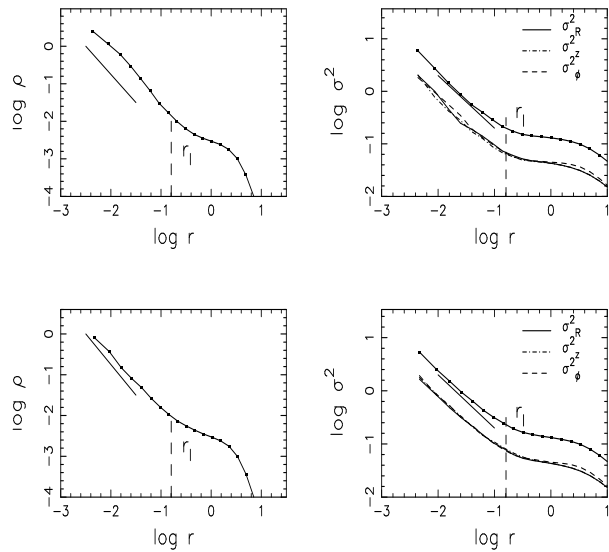

\vbox{
\includegraphics{figures/df1_M002.ps}
\vspace{4.cm}
\includegraphics{figures/df2_M002.ps}
\vspace{4.5cm}
}
\caption{\label{fig:1d_ck0}
Spherically symmetrized profiles for the density and 
velocity dispersions for the standard run. The total velocity
dispersion (thick line) and the dispersions along each direction are displayed.
The sampling distribution used is {\it resp.}
$f_{S1}$ (top) and $f_{S2}$ (bottom).
The positions of the filled squares indicate the grid used;
the lines indicate slopes $\gamma=3/2$,
 for the density, and $\gamma=1$ for $\sigma^2$}
\end{figure}

The logarithmic slopes $\gamma_\rho$ for $\rho(r)$, and $\gamma_\sigma^2$
 for $\sigma^2(r)$, are evaluated by computing the 
respective average logarithmic derivative over the grid points in
some interval $I_r=[0,r_{max}]$.
To estimate the error in this measure, we also compute the slope for
each individual snapshot, and infer the mean deviation.

 The radius $r_{max}$ should in theory be $r_{max} \approx r_I$.
In practice, we take the maximum interval in which the radial profile
appears as a straight line when plotted in logarithmic
axes. 
This leads actually to consider a slightly smaller interval than
$[0,r_I]$ for $\sigma^2(r)$, in order to avoid the transition region
from cusp to core
(see Fig. \ref{fig:1d_ck0}). 
Including grid points that belong to this transition 
region would obviously lead to a systematic under-estimate for the cusp slope.

This point is illustrated in Table \ref{tab:gamma}, for
the standard case with $M_{BH}/M_0=0.02$.
This is the case where 
the measure depends the most on the choice of $I_r$.
Indeed, the less massive the BH, 
the fewer the grid points contained within $r_I$.
 As a consequence, each of the grid points
has a larger statistical weight in the average slope. 

Of course, the value of the slope for $\sigma^2$ must be
$1$ in the region where the BH dominates the dynamics.
As can be seen from the table, the effects of excluding successively 
 $r_N$, $r_{N-1}$ etc.
from the $N$ grid points within $[0,r_I]$ 
are, for $\gamma_{\sigma^2}$: (i) to increase the average slope estimate
(ii) to increase the mean deviation around the average value, as
fewer data points are made available.

If we exclude the $2$ points lying within the cusp/core transition,
the logarithmic slope measured for $\sigma^2$ is very nearly equal to
unity, as it should be.

\begin{table}
\caption{\label{tab:gamma}
The value of the logarithmic slope, estimated for
 respectively $\rho(r)$ and $\sigma^2(r)$; the mean deviation is also indicated.
The case shown is for $M_{BH}/M_0=0.02$, and $f_S=f_{S1}$.
 The successive columns
 correspond to taking different fractions of the total number $N$ of 
grid points within $[0,r_I]$ --see text.}

\hskip 1mm\begin{tabular}{ l   c l l }
\hline
  \hfil   & $n=N =8 $         &  $n=N-1$& $n=N-2$\\
 \hfil &  $(r_{max}=r_I)$ &         &        \\
 $\gamma_{\rho} $    & $1.46 \pm 0.06$ &  $1.43 \pm 0.05$  & $1.43
\pm 0.06$         \\
$\gamma_{\sigma^2}$ &   $0.93 \pm 0.05$  & $0.99\pm 0.06$  & $1.01 \pm
0.08$ \\
\hline 
\end{tabular}

\end{table}

As for the density, there are up to 8 points that lie well within
the cusp (Fig. \ref{fig:1d_ck0}). Our density profile deviates slightly from
a pure power-law at the 2 inner points. Therefore we obtain a better 
estimate of the slope in the linear section by taking into account
the full set of points within $r_I$.
 With these grid points, we find for the
standard case:
\begin{itemize}
\item using $f_{S1}$ :
$\gamma_{\rho} \simeq 1.46 \pm 0.06 $
\end{itemize}

Fig. \ref{fig:1d_ck0} also displays the density and velocity profiles
obtained when the sampling distribution is $f_{S2}$. The central
density value is slightly smaller than with the previous case,
probably because the sampling is not as efficient due to its strong
spherical asymmetry. The value measured for the density logarithmic
slope is now:
\begin{itemize}
\item using $f_{S2}$ :
$\gamma_{\rho} \simeq 1.35\pm 0.15 $, 
\end{itemize}

This value is somewhat smaller than the value for the other sampling, but still 
consistent with it.
 We therefore conclude that the two sampling distributions tested,
in spite of their different behaviour in the centre, yield similar
results. We therefore believe the results are not an artifact of
the method used.
The value of $\gamma_{\rho}$ is consistent, within its error
bars, with the value
given in the literature for the spherical case, $\gamma=3/2$. We 
 find no evidence for a significantly larger, nor smaller, slope.

\begin{table*}
\begin{minipage}[c]{15cm}
\caption{\label{tab:list_runs}
List of experiments with non-rotating models, with
their numerical results. Column 4 gives the logarithmic slope measured
for the density, and column 5 gives that for the total velocity dispersion.
$||M_1||$ gives an estimate of the mass in the
 perturbation. $F_{max}$ is the maximum relative contribution to the total force
 of the self-gravity force. The last column gives an estimate
of the relative error made in evaluating the perturbed quantities
(see \S \ref{ss_PP}). 
}

{\small 
\hbox{
\begin{tabular}{l  c  c  c c c c c}
\hline
$\begin{array}{c} M_{BH} \\ \hline M_0 \end{array}$  &
$\begin{array}{c} M_{BH} \\ \hline M_{core} \end{array}$  &
$\begin{array}{c} T_{grow}\\ {\rm(Yrs)} \end{array}$ &
$\gamma_\rho$&$\gamma_{\sigma^2}$ & $ ||M_1||$ & F$_{max}$ &
$\begin{array}{c} M_1 \\ \hline ||M_1|| \end{array}$  \\
\noalign{\vskip 3mm}
 0.02 ({\it spherical})  & 0.1 & $10^7$ &
$  1.52 \pm 0.02$ &$ 0.94 \pm 0.05$ & $0.05$ &0.01 & $0.08$  \\
 0.02  {\it (standard)} &  0.1
 & $10^7 $ & $  1.46 \pm 0.06 $ & $ 1.01  \pm 0.08 $
& $ 0.05$   & 0.02 & $0.05$ \\
0.03 &  0.15 &  $10^7$ & $ 1.50  \pm 0.04  $
& $  0.98  \pm  0.02 $ & $0.06$   & $0.02$  & 0.03   \\
0.2 &  1 &  $10^7$ & $  1.71 \pm 0.01 $&  $ 0.99 \pm  0.01 $
     & 0.4  & 0.15  & $  - 0.02$ \\
 0.03 & 0.15  & $10^6$ & $1.31 \pm 0.1$ &$ 0.92 \pm 0.06$ 
& $0.07$ & $0.02$ & $0.02$ \\
 0.03 & 0.15  & $10^5$ & $0.92 \pm 0.27$ & $ 0.96 \pm 0.08$ 
& $0.07$ & $0.02$ & $0.02$ \\
\hline

\end{tabular} 
} 
}
\end{minipage}
\end{table*}

The results found for experiments using a larger final BH mass are
discussed in \S \ref{ss_mbh}. For such cases,
 the dispersion in the measures is very small (see Table 
\ref{tab:list_runs}), as the number of points useful for the
measure is larger.

\subsection{Orbital response to the BH}
\label{ss_dfBH}

When we use the PP method, we have direct access to the
distribution functions $f_0,f_1 $ and $f(E,L_z)=f_0+f_1$.
Those are available with a large signal-to-noise ratio, 
even for the case where the BH mass is only $1\%$ of the model mass,
that we analyse in this paragraph.

\begin{figure*}
\begin{minipage}[l]{15cm}
\vskip 10cm
\hskip -2cm
\hbox{
\vbox{\vskip 1cm \includegraphics{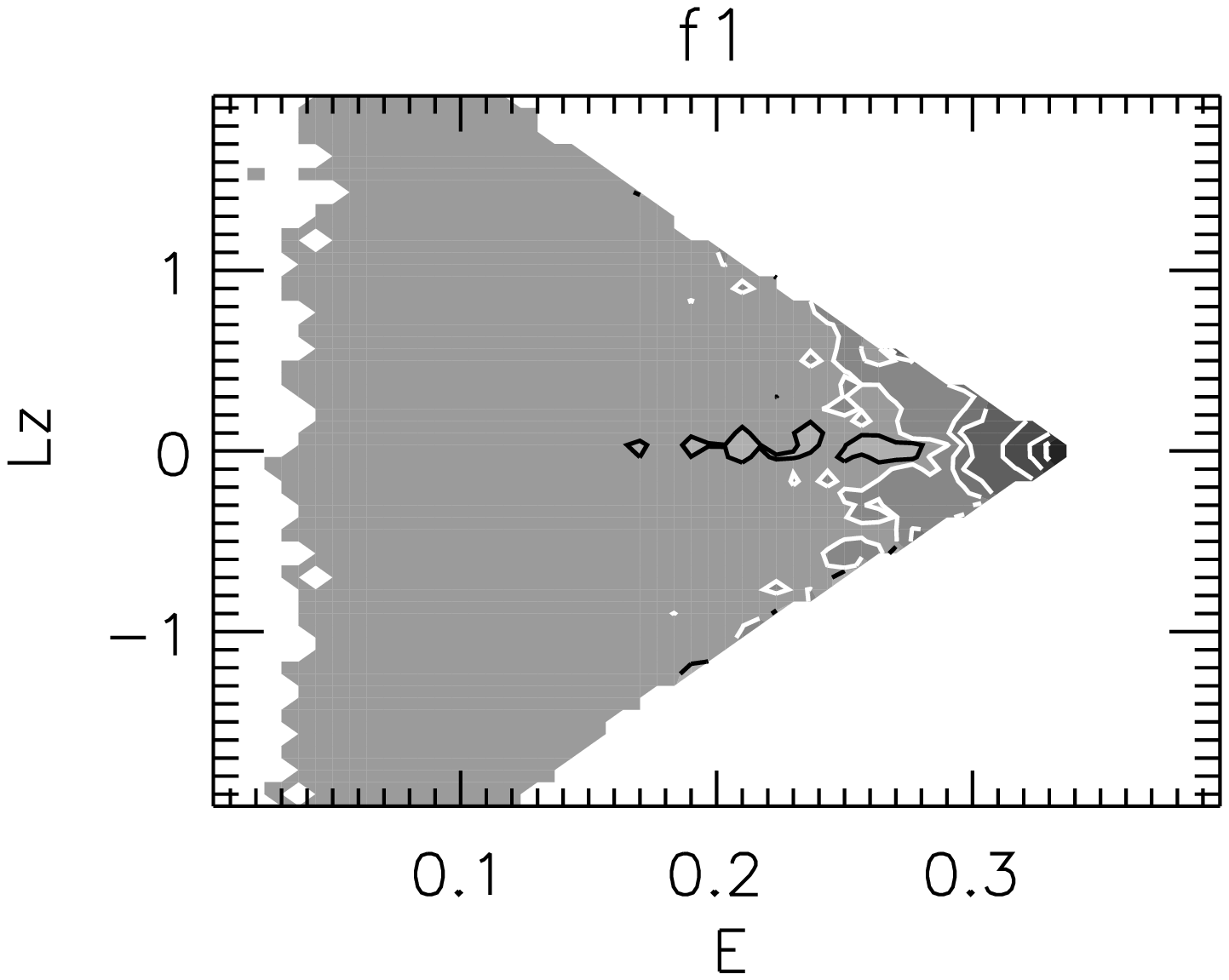} }
 \hskip 5cm \includegraphics{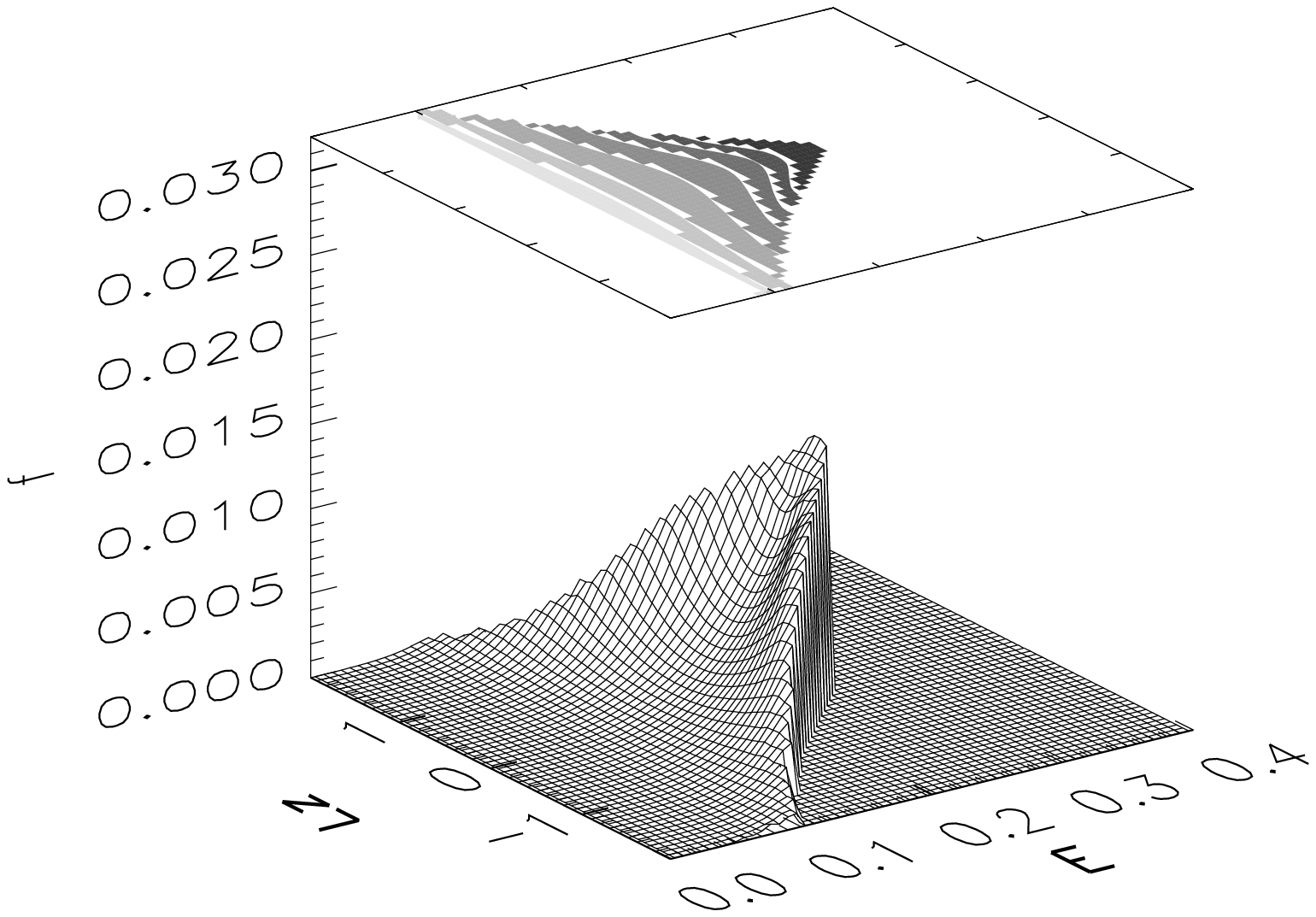}
 \hskip 5.5cm \includegraphics{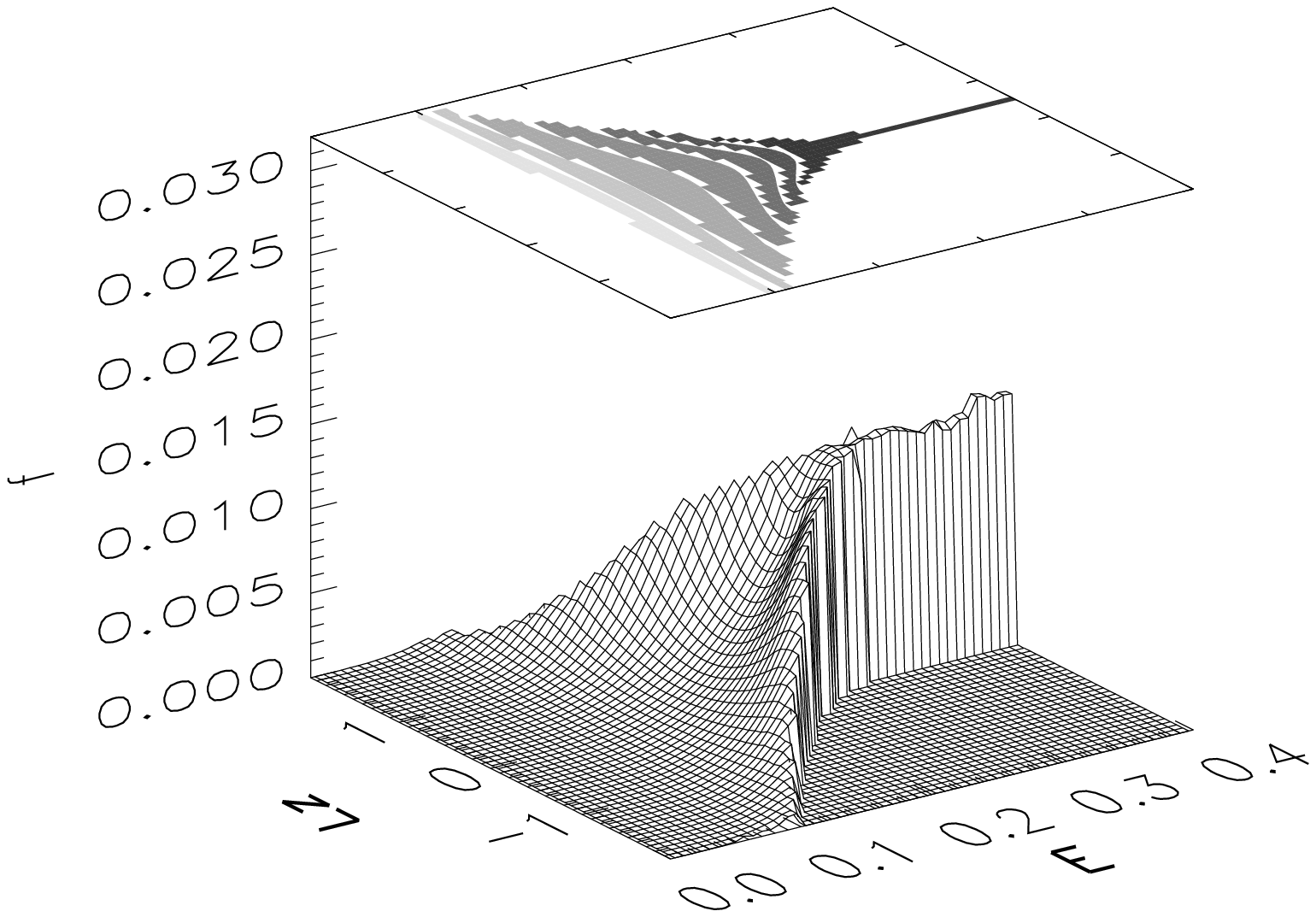}
}
\vspace{5cm}
\vskip -10cm
\caption{\label{fig:df}
Distribution functions for the case without net rotation.
 The left panel shows the perturbation distribution
$f_1(E_0,L_z)$ as a function of the initial unperturbed energy
 and angular momentum of the particles. The grey scale is dark for
positive values, scaling linearly from $10^{-5}$ to $4\times 10^{-3}$
 (with white contours over-imposed), and light (with
black contours) for negative values.
The 3-D plots display the initial $f_0(E_0,L_z)$ (center) 
and final $f(E,L_z)$ 
(right) distribution functions-- see text
 for details on the corresponding $E,L_z$ axes.}
\end{minipage}
\end{figure*}

The initial model can be viewed by plotting the surface
$f_0(E_0,L_z)$, with $E_0 =\Phi_0-{v^2}/{2}$ 
the unperturbed energy. Since $\Phi_0(r=0)=0.25$,
 $0<E_0<0.25$. This surface, drawn from a set of initial
conditions for the particles,
 is shown on Fig. \ref{fig:df} (central panel). We use a regular grid
in $E_0$ and $L_z$ to bin the particles, and sum the
 masses $m_{0i}= f_0(w_i)/f_S(w_i)$ within each cell of the grid. 
It could of course have been drawn analytically, but, as
it is, the figure shows how smooth the numerical realization is.
A map using a linear grey scale is shown on top of the surface. 
At fixed energy, the d.f. increases with increasing $L_z$, as
can be seen from the figure. This is
related to the excess of tangential motion supporting the flattening. 

In a similar way, using snapshots at the end of the standard run,
we can build the surfaces $f_1(E,L_z)$ and $f(E,Lz)$.
We plot in fact two slightly different maps.
(i) A map of $f_1$ as a function of the initial values
of $E_0$ and $L_z$
-- $L_z$ is anyway conserved during the BH growth-- 
reveals which orbits have been most perturbed.
(ii) On the other hand, a map for
$f(E,L_z)= f_0(E,L_z) + f_1(E,L_z)$ (with $E=\Phi_0 +
\Phi_{BH}+\Phi_{s.g.} -\frac{v^2}{2}$ the total
final energy) yields information about 
the orbital structure in the final model.
In practise and for the sake of simplicity, we neglect the self-gravity of the cusp
(we have shown in \S \ref{ss_selfg} that the self-gravity contribution
is small for $M_{BH}/M_0 \leq 0.03$), and 
plot $f(E^*,L_z)$, where $E^*\equiv \Phi_0+ \Phi_{BH} - \frac{v^2}{2}$.

The map for $f_1$ is show on Fig. \ref{fig:df} (left panel).
Positive values of $f_1$ are displayed in grey shades darker
than the background (with white contour lines), whereas negative values appear 
as lighter shades (with black contour lines). 
The map shows that orbits having
$E_0 \simeq \Phi_0$ have been the most perturbed. Also, at
a given energy, the perturbation is more important for orbits with
small angular momentum $|L_z|$.
For the most bound orbits ($E_0 \simeq \Phi_0$), $f_1 >0$,
 reflecting the fact that they now belong
to phase-space regions more populated than initially. By contrast, a negative
perturbation ($f_1<0$) can be found for smaller 
values of $E_0$, and $L_z\simeq 0$.
These orbits have an apocenter located
 at a radius much larger than the influence radius.
The fact that the perturbation extends much further than the
influence radius was already seen in Fig. \ref{fig:dm1} (\S \ref{ss_selfg}).

We have used 10 snapshots, issued at the end of the simulation and
 merged together, in order to produce the surface $f(E^*,L_z)$ shown
on the right panel of Fig. \ref{fig:df}. 
The final total distribution shows most clearly
 that only orbits with $L_z \simeq 0$ have a final energy
$E^*  > \Phi_0 (r=0)$ (Fig. \ref{fig:df}).
 These are orbits such that $\Phi_{BH} - \frac{v^2}{2} > 0$, {\it
i.e.} orbits that have become bound to the BH potential.
Such orbits must provide an important contribution to the cusp.
However, the fact that nearly radial orbits are more efficiently
 attracted into the
cusp, does not imply radially biased motion within the cusp,
since these orbits get rounder (see also CB94). 
In models with a central cusp, constructed either
 with, or without, a BH, Dehnen (1995) and 
Dehnen \& Gerhard (1994) find that such orbits  
typically have $L_z \simeq L_c(E)$.

The final d.f. appears remarkably
 constant at high binding energies ($ f(E)\simeq cst$ for $E >
 \Phi_0(0)$ and $Lz \simeq 0$).
 Tremaine {\it et al.} (1994)
have built analytical models for spherical systems 
with a central density cusp and a central
 BH. For a cusp with slope $\gamma=3/2$, their d.f. tends to a
constant for high binding energies, in a way very similar to our results.

The final d.f. can therefore be said `degenerate' at high binding
energies, in the sense
that different energy levels have the same probability.
Such a degeneracy is also typical of
violently relaxed systems (see Lynden-Bell 1967, Chavanis \&
 Sommeira 1998). 
Our results therefore support those by 
Stiavelli (1998), who has shown that the observable signatures
 of an adiabatically grown BH, on the one hand, 
and of violent relaxation around a pre-existing BH, on the other hand,
 are very similar. We have shown that this is due to the fact
 the distribution functions themselves have very similar behaviour at
 high energies.

\subsection{Influence of growth time}
\label{ss_tbh}

The influence of the BH growth time on the final cusp
has been checked,  
 by experimenting with shorter times ($T_{BH} =10^6$, $10^5$ Yrs), for
the case $M_{BH}/M_0=0.03$.
The dynamical times within the cusp are 
close to $10^5$ Yrs, therefore the
assumption of adiabaticity does not hold for the latter case.
Moreover, some orbits may not have had time
to rearrange into an equilibrium during the BH growth. Therefore all 
simulations are pursued after the BH has reached its final mass, so
that the total time is $10^7$ Yrs for all cases. 

Results for the logarithmic slopes are reported in Table \ref{tab:list_runs}.
In good agreement with Sigurdsson {\it et al.} (1995), we 
found that the adiabatic model is still roughly valid for growth times
of order $\sim 10 \, T_{dyn}$.
Nevertheless small growth times introduce some differences, and
already for $T_{BH}=10^6$ Yrs, there is a tendency for the cusp
to spread over a larger diameter. Whereas the density profile follows a 
power-law in a smaller region around the center, 
 the transition zone around $r_I$ is
much larger than in the case $T_{BH}=10^7$ Yrs.
There is therefore a trend for the average cusp slope to 
be slightly smaller than $3/2$.

\begin{figure}
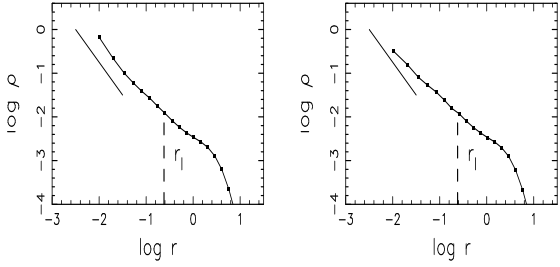

\hbox{
\includegraphics{figures/T6_M003.ps}
\hskip 3.8cm
\includegraphics{figures/T5_M003.ps}
}
\vspace{5 cm}
\caption{ Density profile for a model with final $M_{BH}/M_0=0.03$ 
and with BH growth time respectively $T_{BH}=10^6$ Yrs (left) and
$T_{BH}=10^5$ Yrs (right). The line segments indicate the $\gamma=3/2$
power-law.}
\label{fig:T5_M003}
\end{figure}

This trend towards a shallower cusp is more evident in the
case where $T_{BH}=10^5$ Yrs (Fig. \ref{fig:T5_M003}).
Actions are not expected to be conserved in this case where $T_{grow} 
 \sim T_{dyn}$, as orbits are deflected by the rapidly
varying central potential. Probably as a result of such deflections, 
we find an excess of radial velocity 
dispersion ($\sigma_R \simeq \sigma_z > \sigma_{\phi}$) in the central
region, as found also by Sigurdsson {\it et al.} (1995) for
experiments with a similar growth time. Less orbits settle 
in the vicinity of the BH, and the cusp appears less strong
than in the adiabatic case, with a logarithmic slope $\gamma_\rho
\simeq 0.9$. 

\subsection{Varying the BH mass}
\label{ss_mbh}

We also experiment with BH masses which are a larger fraction
of the model mass (see Table \ref{tab:list_runs}).
The mass of the dark matter concentrations betrayed by 
a central cusp in early-type galaxies is
evaluated to be at most $2 \%$ of the
galaxy mass ({\it e.g.} Richstone {\it et al.} 1998). Maybe more 
significant for the cusp morphology and kinematics, is the 
ratio between the BH mass and the initial core
mass of the galaxy.

Semi-analytic works (Young 1980, Quinlan {\it et al.} 1995)
predict that the density profile of the cusp remains roughly
self-similar, as long as the final values of the BH mass  
are smaller than the mass within the initial core.
On the other hand, for $M_{BH}$ larger than the core mass, 
the cusp is steeper, with $\gamma_\rho$ approaching a value
of $2$ when the mass ratio is $M_{BH}/M_{core} \gg 1$.
 Of course no variation is seen in the velocity
dispersion cusp, which is always what we expect in an essentially 
keplerian potential ($\gamma_{\sigma^2} =1$).

\begin{figure}
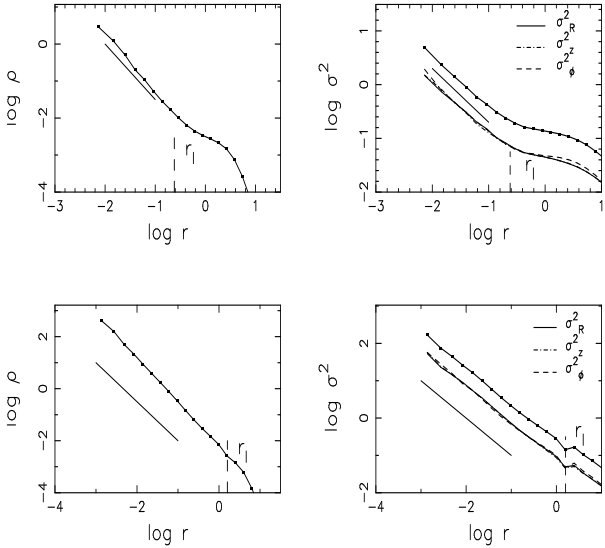

\vbox{
\includegraphics{figures/df1_M003.ps}
\vspace{4.cm}
\includegraphics{figures/df1_M02.ps}
\vspace{4.5cm}
}
\caption{Spherically symmetrized profiles for $\rho(r)$ and
$\sigma^2(r)= 1/3(\sigma^2_R +\sigma^2_z + \sigma^2_\phi)$, shown
for $M_{BH}/M_{core}= 0.15$ (top panels) and $M_{BH}/M_{core}=1$
(bottom panels). The dispersions along
individual directions $\sigma^2_r$, $\sigma^2_R$ and $\sigma^2_\phi$
have a shape very similar to that of the total velocity dispersion.
}
\label{fig:1d_ck0_M}
\end{figure}

The cases reported in Table \ref{tab:list_runs} correspond to
a BH with mass equal respectively to $0.05$, $0.15$ and $1$ times
 the core mass $M_{core}$. The first case has already been discussed in
\S \ref{ss_slope}; the two other cases are illustrated on Fig. 
\ref{fig:1d_ck0_M}, which 
displays the spherically symmetrized profiles for the density
and the velocity dispersions.

As can be seen from this Figure, the number of points within
the power-law cusp is much larger than for the standard case.
For a BH whose mass is a substantial fraction of the core mass,
the statistics on the results is therefore much
better than in the standard case $M_{BH}/M_{core}=0.1$. 
 
\begin{figure}
\includegraphics{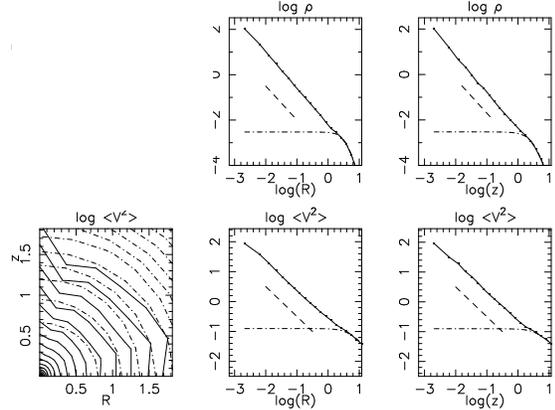}
\vspace{7cm}
\caption{Left: Isocontours for the final density, and profiles along
$R$ and $z$, for $M_{BH}/M_0=0.2$.
Isocontours and profiles for the velocity dispersion are displayed
on the second line.}
\label{fig:maps02}
\end{figure}

A single snapshot (with $N\simeq 320,000$ particles)
suffices to trace clearly the cusp, and measure its slope
with good accuracy. The dispersion in this measure
is very small, as can be seen in the results 
summarized in Table \ref{tab:list_runs}. 
Fig. \ref{fig:maps02} shows in more detail the case 
with $M_{BH}= 0.2 \, M_0$,
corresponding to $M_{BH}/M_{core} \simeq 1$. 
It was plotted using one snapshot.

\begin{figure}
\includegraphics{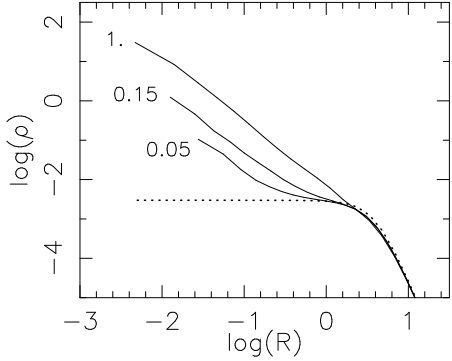}
\vspace{6.5cm}
\caption{Density profile along $r$ obtained for a BH with
final mass {\it resp.} $0.05, 0.15, 1.$ times the
initial core mass}
\label{fig:profil}
\end{figure}

The logarithmic slope $\gamma_{\rho}$ that we measure for the
total density is an increasing function of the final BH mass (see
Table \ref{tab:list_runs}). 
It takes values within $[3/2, 2]$, very much as
expected from the spherical adiabatic model. 
Fig. \ref{fig:profil} compares the density profiles obtained 
for the different $M_{BH}/M_{core}$ values considered.

We also ran a simulation where the final BH mass was $2\, M_{core}$.
Although results seemed very reasonable in all the cases considered,
 the error in the mass conservation increases in this case to
non-negligible values ($M_1/||M_1||\sim 20 \%$). The computation 
 is therefore at the limit of credibility for the method.
The perturbation technique used here
is indeed better suited to small or moderate values
of $M_{BH}/M_0$, rather than larger ones. The reason is not
that the method is limited to the linear regime: 
the computation is fully non-linear, since we compute
perturbed orbits within the full potential. The reason
is rather that the sampling is more
difficult to control for larger perturbations. The non-conservation of
the total mass reflects the sampling inadequacies.
For higher mass ratios than those considered in this work,
a standard $N$-body technique should probably be preferred.

\section{Cusps in rotating systems}
\label{sec:cusprot}

In this section, we investigate the effects of an initial rotation of
the host galaxy. LG89 and CB94 both study the rotation brought to
central regions by a growing cusp, but find different results.
 Is the galaxy response somehow affected in these works by 
the approximation of a spherical potential?
Here we can study this case in its true geometry, without
any such approximation. Moreover, we would like to clarify which
difference in these works is responsible
for the difference in the final velocity field.
The net rotation within the influence radius
is weak in CB94, unless the
BH has a mass much larger than the initial core mass.
On the other hand, LG89 obtain a cusp in the rotation
velocity for any initial BH mass. We therefore investigate in turn models that
are built similarly to those considered by CB94, and LG89. 

\begin{table*}

\centering{
\begin{minipage}{15cm}
\caption{\label{tab:listrot}
List of numerical experiments with rotating models. See 
Table \ref{tab:list_runs} for the symbols definition. The results are
shown for $p=p_1(E,L_z)$; experiments with $p=p_2(E,L_z)$ yield
practically identical measures for the logarithmic slopes 
of both $\rho$ and $\sigma^2$.}

{\small 
\hbox{
\begin{tabular}{c  c  c c c c c c}
\hline
$\begin{array}{c} M_{BH} \\ \hline M_0 \end{array}$  &
$\begin{array}{c} M_{BH} \\ \hline M_{core} \end{array}$  & 
$\begin{array}{c} T_{grow}\\ {\rm(Yrs)} \end{array}$ &
$\gamma_\rho$&$\gamma_{\sigma^2}$ & $||M_1||$ & F$_{max}$ &
$\begin{array}{c} M_1 \\ \hline ||M_1|| \end{array}$  \\
 0.01  &  0.05 &
  $10^7 $ & $  1.48 \pm 0.1 $ & $  0.9 \pm  0.1 $
& $ 0.02$    & 0.01 &  $0.02$ \\
0.03 &  0.15 &  $10^7$ & $ 1.53 \pm 0.04  $
& $ 0.92  \pm 0.1 $ & $0.05$   & $0.02$  & 0.03   \\
0.2 &  1 & $10^7$ & $ 1.66 \pm 0.04  $
& $   1.01 \pm 0.03 $   & 0.4  & 0.08  &$ - 0.02$ \\
\hline
\end{tabular} 
} 
}
\end{minipage}
}
\end{table*}

\subsection{Models with $p_1(L_z)$}
\label{ss_p1}

We first experiment with rotating models built using $p(E,L_z) =
p_1(L_z)$ (Eq. \ref{eq:defp}).
The motivation for considering this kind of
rotation tapering for small $L_z$, is obviously its simplicity.
Furthermore, such models are not unplausible
if a galaxy has formed through a major merger. Violent
relaxation in the central regions can erase net rotation
for very bound orbits, whereas less bound orbits may still exhibit
a preference for one sense of rotation. An example of merger remnant
with this sort of angular momentum distribution is found in numerical
 simulations by Barnes \& Hernquist (1996; their Fig. 17).

The parameters in $p_1$ have been chosen as follows:
\begin{equation}
  L_m = 0.2 \, l_0   ~~~~~~~~ \eta=0.5.
\label{eq:lm1}
\end{equation}

\begin{figure}
\vskip 7cm
\includegraphics{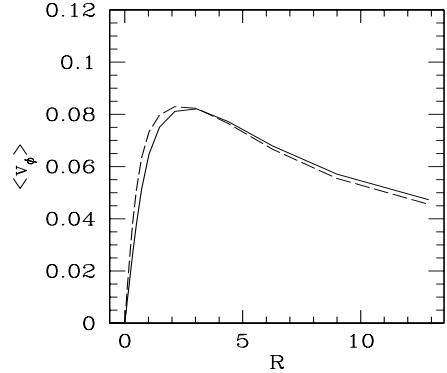}
\vspace{6cm}
\vskip -7cm
\caption{Initial rotation velocity profile along the $R$ axis, for the model
with {\it resp.} $p=p_1$ (solid line), and $p=p_2$ (dashed line), as
given by Eq. \ref{eq:defp}}
\label{fig:ctan0}
\end{figure}

\noindent
The corresponding rotation curve is displayed on Fig. \ref{fig:ctan0}.

For each BH mass considered, the slope of the cusp is
the same, within the error bars, as in the
corresponding non-rotating case. These results are summarized in 
Table \ref{tab:listrot}. 

\begin{figure*}
\begin{minipage}[l]{15.5cm}
\vskip 10cm
\hbox{\hskip -2cm
\vbox{\vskip 1cm
\includegraphics{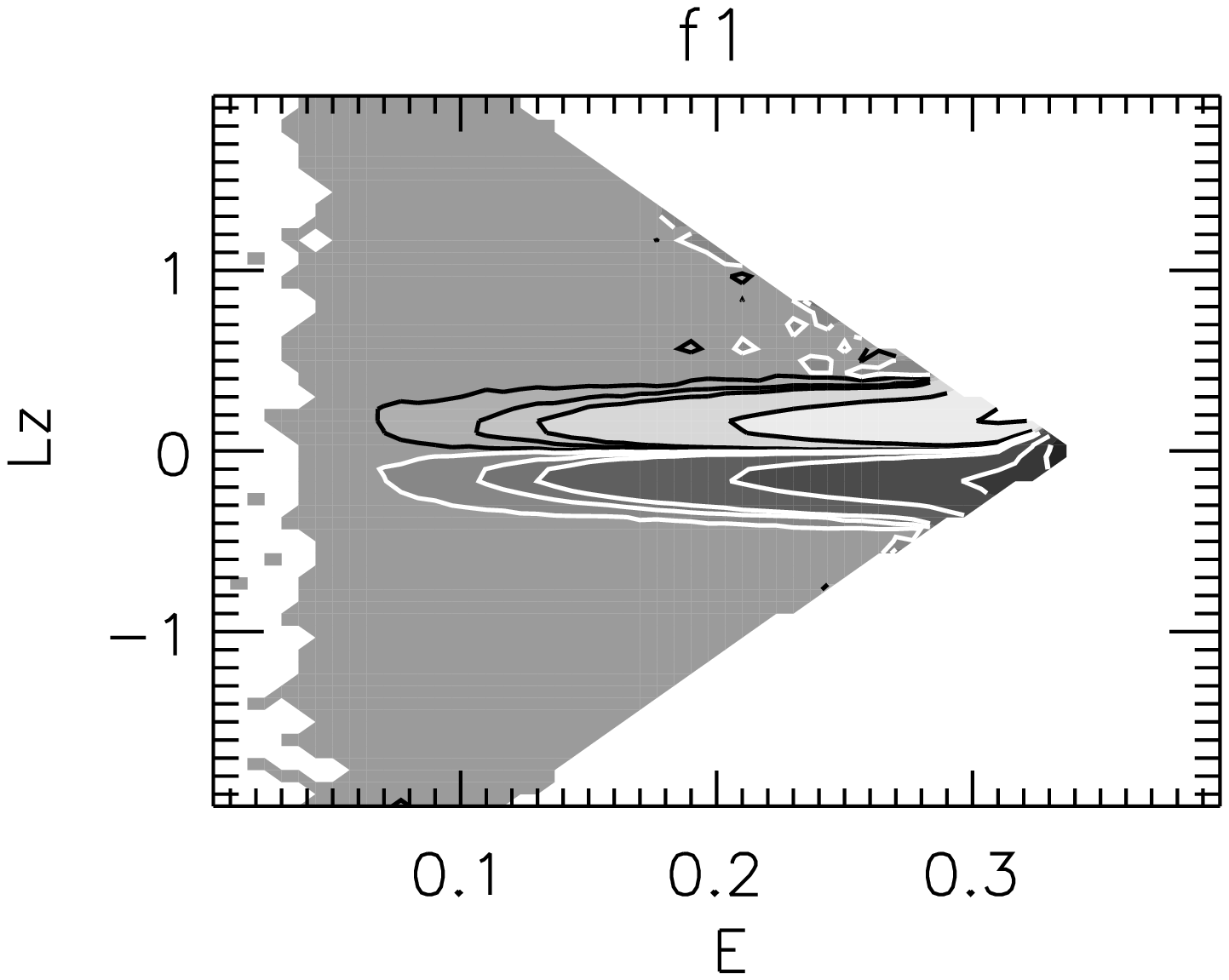}}
\hskip 5cm
\includegraphics{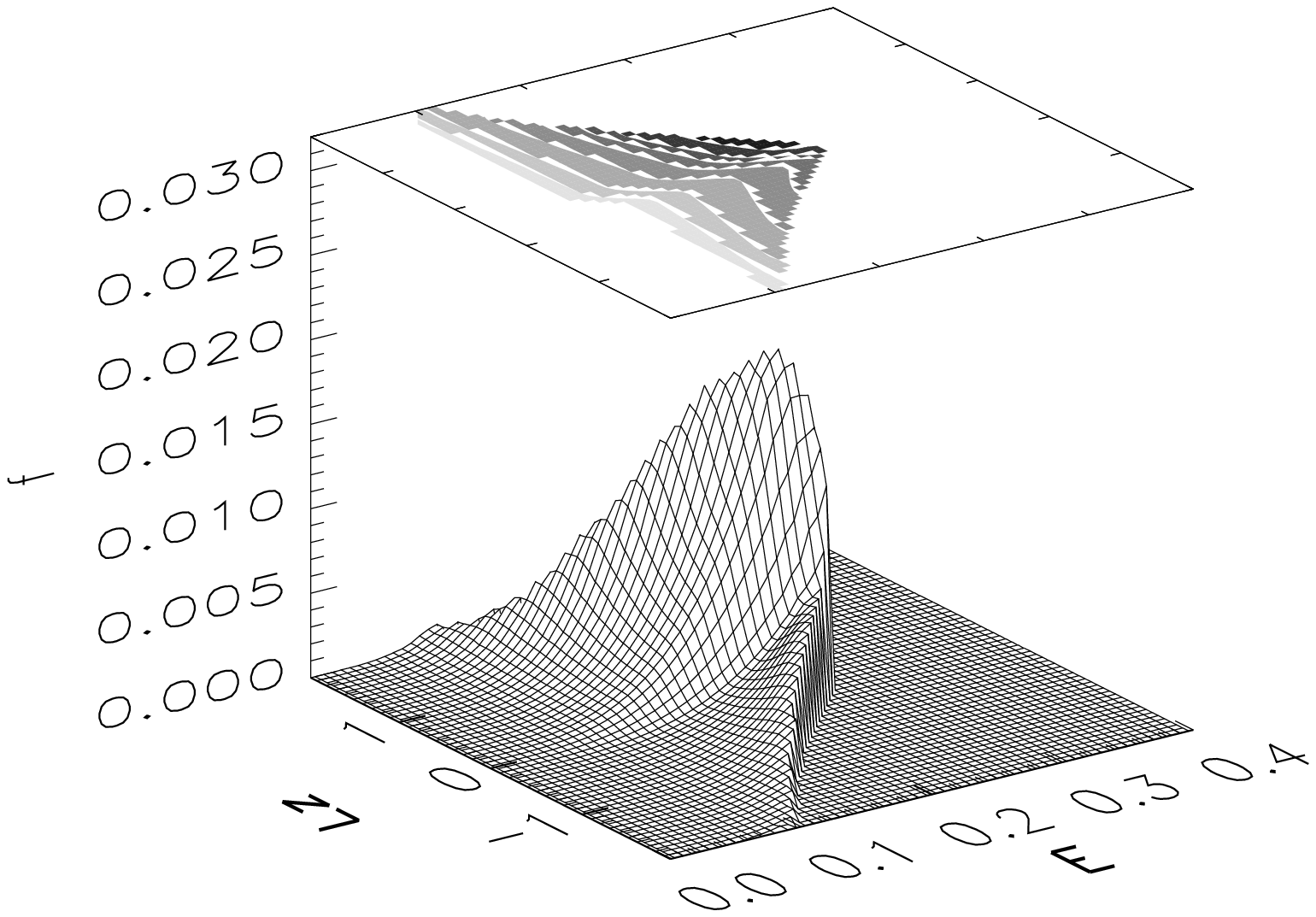}
\hskip 5.5cm
\includegraphics{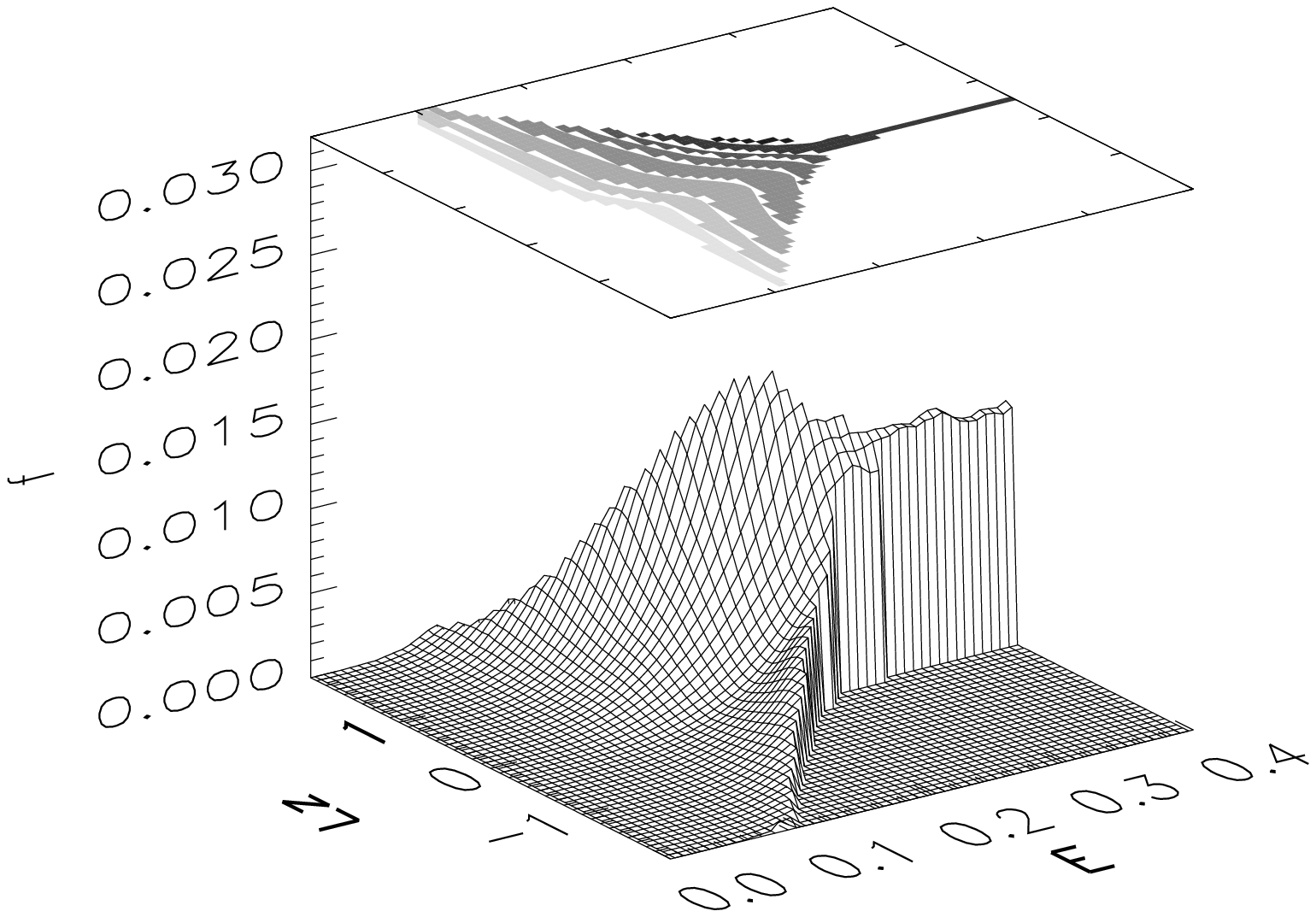}
}
\vspace{5cm}
\vskip -10cm
\caption{Distribution functions for the case $M_{BH}/M_0=0.01$ with rotation,
and $p=p_1$. The three panels are as in Fig. \ref{fig:df}, displaying
 respectively $f_1(E_0,L_z)$ (left), $f_0(E_0,L_z)$ (middle) and
$f(E,L_z)$(right). The grey
scales and contour lines used for the left panel are defined as in Fig.
\ref{fig:df}}
\label{fig:dfrot}
\end{minipage}
\end{figure*}

The initial distribution $f_0(E_0,L_z)$ is displayed
on the middle panel of Fig. \ref{fig:dfrot}, for a mass ratio $M_{BH}/
M_0 =0.01$, corresponding to $M_{BH}/M_{core} \simeq 0.05$.
 Again, even for this small final BH
mass, the distribution function exhibits little noise.
 Preference for the positive
sense of rotation is reflected by the distortion of the
surface $f_0(E_0,L_z)$, which is higher for $L_z >0$ values.

The left panel of Fig. \ref{fig:dfrot}  shows the perturbation
$f_1$ as a function of the initial integrals of motion $E_0, L_z$.
The grey shades and contour lines are as in the left panel of
Fig. \ref{fig:df}. Again, perturbation is strongest for large $E_0$ values,
and small $L_z$ values. The peak of positive perturbation (at $E\simeq E_0$)
is roughly even in $L_z$, corresponding to an energy range where 
the initial distribution $f_0$ is itself roughly even in $L_z$.
We have indeed considered a model like the model in CB94,
where the net rotation is mainly due to orbits having
$L_z \ge L_0$, with $L_0$ some finite value. For orbits with
high binding energy, $L_c(E)$ is smaller than $L_0$, so that
practically none of them can contribute to the net rotation. 
On the other hand, for smaller energies $E_0$, the perturbation $f_1$ is odd with
respect to $L_z$. This is a consequence of the odd
term $\eta p(L_z)$ that has been added to the distribution function
$f_0$ in order to construct $f_0^r$ -- see Eq.
\ref{eq:defrot}. 

The final total distribution function $f(E,L_z)$ is displayed on the
same figure (right panel). At high energies, the distortion of this
surface towards positive $L_z$ has slightly
decreased in amplitude, corresponding to the negative perturbation
$f_1$ for the positive $L_z$. 
The strip of particles having $E>\Phi_0(0)$
is very narrow around $L_z=0$, as can be best seen on the
grey scale map on top of the right panel. Thus,
 only orbits with $L_z \simeq 0$ become bound to the cusp.
As a consequence, orbits confined to the central regions
do not create a significant global rotation.

\begin{figure}
\vskip -0.5cm
\includegraphics{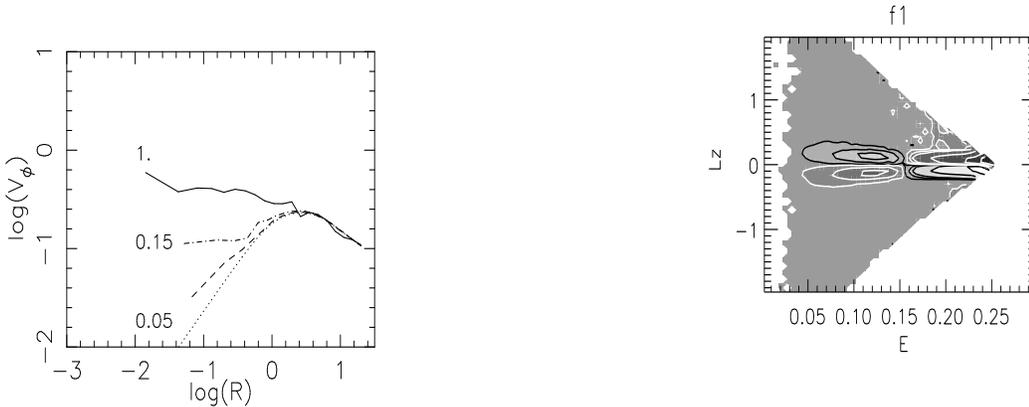}
\vspace{6.5cm}
\caption{Rotation velocity profiles, along the $R$ axis, 
for three different mass ratios $M_{BH}/M_{core}= 0.05, 0.15, 1$, 
as labeled; the dotted line shows the initial rotation of the model}
\label{fig:vrot}
\end{figure}

 Fig. \ref{fig:vrot} shows 
the rotation at the end of the BH growth for different mass ratios. 
Similar to results by CB94, we find that a moderate BH mass does
not bring a significant rotation within the influence radius. 
This result therefore was not affected
by simplifications in the analytical derivation of CB94.
Our results are indeed very close to those of CB94, as
can be seen by comparing our Fig. \ref{fig:vrot} and their Fig. 3.

As the BH mass increases well above the core mass, its influence radius
eventually overcomes the radius corresponding to the circular
orbit with angular momentum $L_0$, and rotation
becomes more important in the centre.
However, for the mass ratios we consider (up to
 $M_{BH}/M_{core} \sim 1$),
we find that no cusp is produced in the rotation velocity. 

\subsection{Models with $p_2(E,L_z)$}
\label{ss_p2}

We now experiment with a rotating model that uses
$p(E,L_z)=p_2(L_z/L_c(E))$
given by Eq. \ref{eq:defp}.
 This model is similar to that by LG89, with their parameter 
$\omega$ corresponding roughly to $\pi/(2 x)$ in our models. 
We obtain a model with an initial rotation very close to the
 rotation for the models with $p_1$ previously considered, by taking:
\begin{equation}
  x= 0.2,  ~~~~~~~ \eta=0.5.
\label{eq:lm2}
\end{equation} 
The corresponding rotation profile along the $R$ axis can be seen 
 on Fig. \ref{fig:ctan0}.

\begin{figure}
\vskip 10cm \hskip -0.5cm
\includegraphics{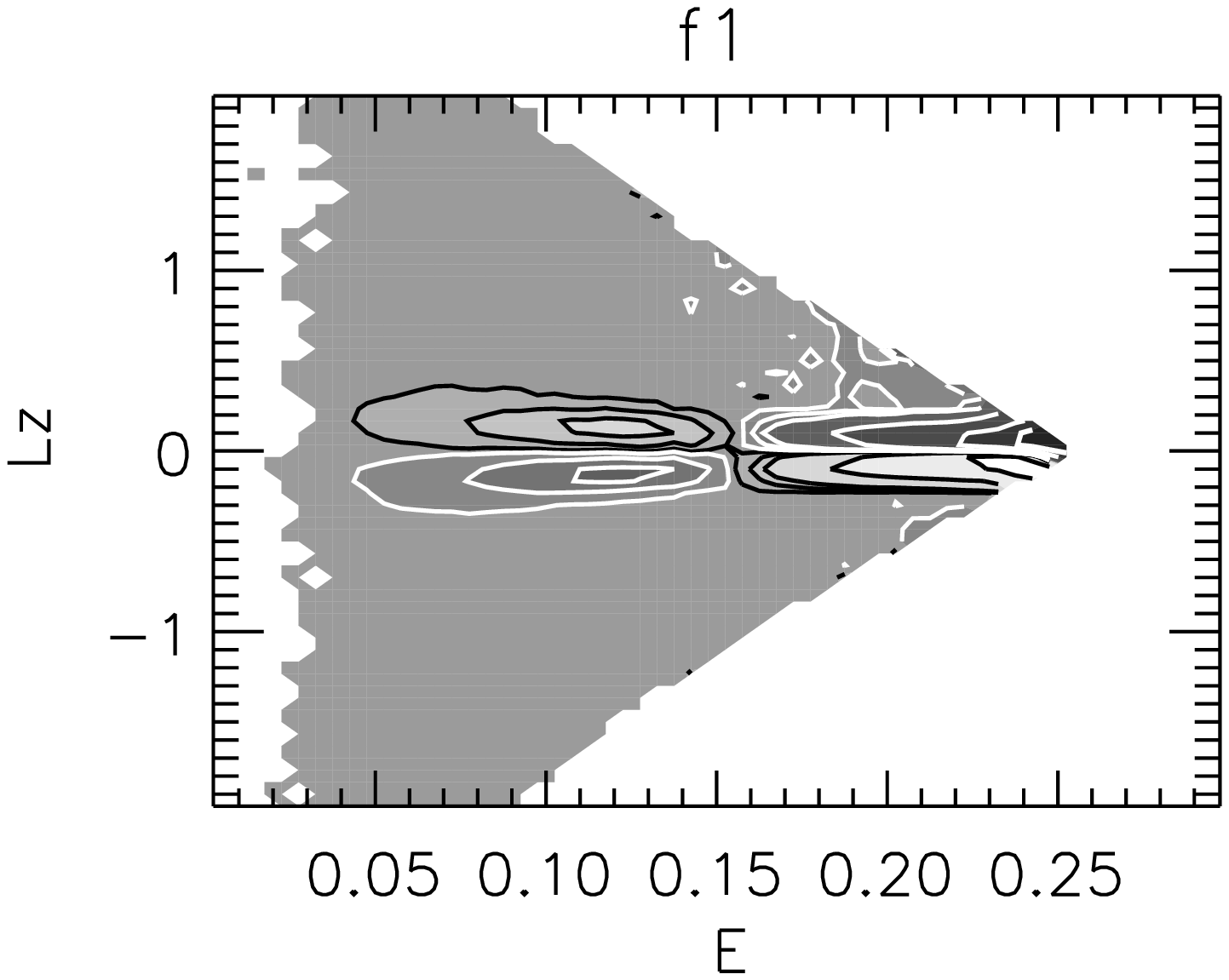}
\vskip -10cm
\vspace{5cm}
\caption{Perturbation distribution function $f_1$ as a function
of the initial energy $E_0$ and $L_z$, for the case  $M_{BH}/M_0=0.01$
with rotation,
and $p=p_2$.}
\label{fig:f1_rot2}
\end{figure}

In the models now considered, an identical fraction of orbits,
 at any energy level,
contributes to the net rotation. Orbits having initially
$E \approx \Phi_0 (0)$ have $L_c(E) \approx 0$,
so that even orbits with very small values of $L_z$ contribute to the net
rotation. These orbits are those which will bring rotation within the cusp.
 A map for $f_1(E_0(t=0),L_z)$
(see \S \ref{ss_dfBH}), displayed on Fig. \ref{fig:f1_rot2},
 shows indeed that, for orbits with high initial
energies, the perturbation has grown positive for $L_z>0$, and
negative for $L_z<0$. Therefore a rotational velocity has efficiently
built within the cusp.

\begin{figure}
\vskip 11.5cm
\includegraphics{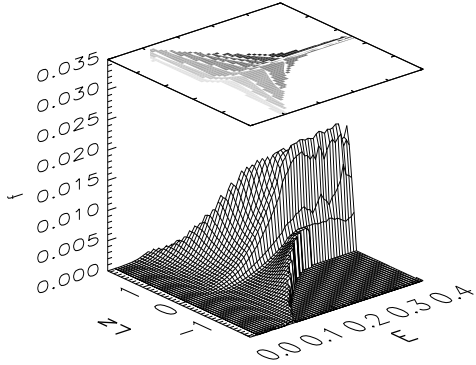}
\vspace{6.5cm}
\vskip -12cm
\caption{Final total distribution functions for the case with rotation,
and $p=p_2$; the case with $M_{BH}/M_0=0.03$ is shown, for greater 
legibility.}
\label{fig:dfrot2}
\end{figure}

The effects on the final distribution function
are best seen from Fig. \ref{fig:dfrot2}. 
We have plotted the case with$M_{BH}/M_0=0.03$, rather than the
 standard case with $M_{BH}/M_0=0.01$, because the structure within
the total final d.f. is more easily visible.
The fact that rotation has been dragged within the density
cusp is disclosed by the bent of the surface strip of 
 particles with high binding energies.

\begin{figure}
\includegraphics{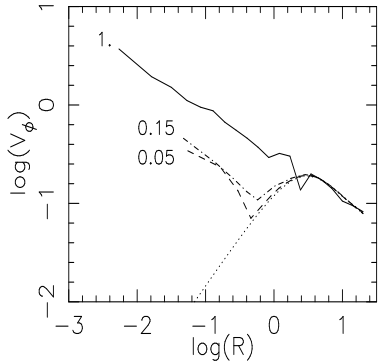}
\vspace{6.5cm}
\caption{Same as Fig. \ref{fig:vrot},
but for an initial model with $p(E,L_z)=p_2$ -- see text.
The label next to each curve refers to the mass ratio $M_{BH}/M_{core}$.}
\label{fig:vrot2}
\end{figure}

This explains why a cusp in rotation velocity is now produced for any BH mass
 considered, in agreement with the models described by LG89.
The cusp follows from the fact that, initially, rotation was present at all
 levels of energy, including high energies.
The rotation velocity profiles along the $R$ axis are shown
 on Fig. \ref{fig:vrot2}, for different masses of the BH.
The different curves are labeled as in Fig. \ref{fig:vrot}.

Therefore we have shown that the rotation in the cusp region
 depends on the orbital structure that existed in the galaxy prior
 to the BH growth. 
The formation of a cusp in $V_{rot}$ around a BH 
that grows adiabatically is sensitive to the way the initial d.f.
depends on $L_z$.
If rotation is not present initially at high binding energies,
only very massive BHs --with respect to the initial core mass -- 
will produce an observable signature
onto the $V_{rot}$ profile.

There has been some debate about the role played by a nuclear disk
onto the light profile in the center of E galaxies (see Jaffe {\it et
 al.} 1994, and Lauer {\it et al.} 1995).
Such a disk would of course demand a different model for
the rotation profile. 
The suggestion that the power-law galaxies (those with steep
 cusps) owed their inner luminosity profile to the presence of
a disk has been much weakened by the fact that Lauer \etal (1995) find little 
evidence for inner disks in the power-law E's of their sample.
More evidence for central disks, at the scale of $\sim 10 $ pc, 
has been collected for SO galaxies (van den Bosch \etal 1994,
 Scorza \& van den Bosch 1998), or giant E's with shallow cusps
(Forbes 1994). Obviously the statistics on the detection of such
 disks is still too small, and the dynamical models far from entering 
 such detail.

\section{Conclusion}
\label{sec:conclu}
This paper aims at investigating the cusp induced in elliptical 
galaxies by the growth of a central, supermassive black hole. 
 The spherical case is the only one to have been studied in detail
in the literature previously, using an adiabatic model.
In this paper we study the cusp due to a growing BH by numerical
means, in order to free ourselves from the assumption of spherical symmetry.

We thus investigate the possible influence
 of flattening and rotation, by considering a simple axisymmetric model. 
For our models, which have a substantial central flattening, we do not find 
any sizeable signature on the cusp slope with respect to the spherical
case. We deduce from this result that a reasonable degree of tangential
anisotropy in the stellar velocities, which sustains the flattening,
 has little effect on the cusp slope. The spherical adiabatic
models appear very robust with respect to the geometry of the initial galaxy.
We find a cusp slope of $1.5$ in the models where the BH mass is less than
the initial core mass, whereas the observed preferred value
for low mass E's is larger, around $1.9$. This fact remains to be
explained. One possibility is that the BH grew in a galaxy with
 initially a very small core,
 as suggested by van der Marel (1999), or even no core at all.
 The observational trends (steeper cusps for low luminosity
E galaxies, and smaller cores for less massive core Ellipticals)
are roughly consistent with this suggestion (see van der Marel 1999). 
Violent relaxation during a merger or a gravitationnal collapse
favours highly concentrated systems
(Farouki {\it et al.} 1983;
see also Chavanis \& Sommeira 1998), specially
if dissipation occurs (Udry 1993). Gas or stellar collisions around the BH 
may also have favoured steeper cusps in low $L$ galaxies. The relaxation time
is indeed shorter for less massive E galaxies as they have denser centers
({\it cf.} Magorrian \& Tremaine 1999)
and we know that the cusp induced in a
collisional system is higher than in the collisionless models, 
with $\gamma_\rho$ around $7/4$ ({\it cf.} Duncan \& Shapiro
1983). 

The formation of a cusp in $V_{rot}$ around a BH 
that grows adiabatically appears sensitive to the way the initial d.f.
depends on $L_z$.
If rotation is not present initially at high binding energies,
only very massive BHs will produce a signature
in the $V_{rot}$ profile. Such an initial orbital
distribution could for instance be expected after mergers involving
an efficient violent relaxation in the central region. 
Stellar kinematic observations with the required resolution
are extremely scarce, although STIS observations ({\it e.g.} Joseph
{\it et al.} 2000)
 may soon improve on this situation.
Existing data for M32 (Joseph {\it et al.} 2000) and for
the E5 galaxy NGC3377  (Kormendy {\it et al.} 1998) show
 an increase of rotation velocity within the 
inner few arcsec.
 This is again consistent with the interpretation that in power-law
elliptical galaxies the BH mass exceeds the initial core mass (eventually zero).

 On the other hand, a cusp in rotation is produced,
whatever the BH/core mass ratio, when initially the fraction of
orbits contributing to rotation is non-zero at high binding energies.
Evidence for such a cusp is probably present in
HST observations of the SO galaxy {\sl NGC} 4342 (Cretton \& van de Bosch
 1999).

Let us finally note that the spherical adiabatic model yields values for the
mass of dark matter concentration similar to more elaborate
dynamical modeling (van der Marel 1999). This however does not
preclude that central BHs 
pre-existed to their host galaxy formation, since 
the adiabatic growth scenario predicts final galaxy models very similar
to those produced by
 violent relaxation around a pre-existing BH (as demonstrated by
Stiavelli 1998, and by our own results on the distribution function).
\vskip 1cm

\noindent
{\large \bf Acknowledgments}

 We are grateful to J.-C. Lambert for his extremely
 efficient and cheerful help with implementing the code on the GRAPE
 machines. We thank P.-H. Chavanis and A. Bosma for stimulating discussions. 
F.L. acknowledges gratefully a Marie-Curie fellowship, as well as
hospitality at the Marseille observatory, during this work. 
We would also like to thank IGRAP, the
INSU/CNRS and the University of Aix-Marseille I for funds to develop
the computing facilities used for the calculations in this paper.

\vskip 0.5cm

\noindent
{\large{\bf References}}

\vskip 2mm
\parskip=0pt
\noindent
Athanassoula, E., Bosma, A., Lambert, J.-C., Makino, J.,  1998 {\it MNRAS}~ {\bf 293},
369

\noindent
Barnes, J., Hernquist, L. 1996, {\it ApJ}~ {\bf 471}, 115

\noindent
Bender, R. {\it et al.} 1989, {\it A\&A}~ {\bf 217}, 35

\noindent Binney, J., Tremaine, S. 1987. {\it `Galactic Dynamics'},
Princeton Univ. Press, Princeton, NJ

\noindent
Chavanis, P.-H., Sommeira, J. 1998, {\it MNRAS}~ {\bf 296}, 569

\noindent
Cipollina, M., Bertin, G. 1994, {\it A\&A}~ {\bf 288}, 43  (CB94)

\noindent
Cretton, N., van den Bosch, F. C. 1999, {\it ApJ}~ {\bf 514}, 704
 
\noindent
Dejonghe, H. 1986,{\it Phys. Rep.}, {\bf 133}, 218 

\noindent
Dejonghe, H., de Zeeuw, T. 1988, {\it ApJ}~ {\bf 329}, 720


\noindent
Dehnen, W. 1995, {\it MNRAS}~ {\bf 274}, 919

\noindent
Dehnen, W., Gerhard, O. 1994, {\it MNRAS}~ {\bf 268}, 1019-1032.

\noindent
Duncan, M. J., Shapiro, S. L. 1983, {\it Nature}~ {\bf 262}, 743

\noindent 
Duncan, M. J., Levison, H. F., Lee, M.-H. 1998, {\it  AJ}~ {\bf 116}, 51 

\noindent
Evans, N. W. 1994, {\it MNRAS}~ {\bf 267}, 333-360.

\noindent
Farouki, R., Shapiro, S., Duncan, M. 1983, {\it ApJ}~ {\bf 265}, 597

\noindent 
Forbes, D. A. 1994, {\it AJ}~ {\bf 107}, 2017

\noindent
Gebhardt, K. {\it et al.} 1996, {\it AJ}~ {\bf 112 }, 105

\noindent
Gerhard, 0. \& Binney, J. 1985, {\it MNRAS}~ {\bf 216}, 467

\noindent
Goodman, J., Binney, J. 1984, {\it MNRAS}~ {\bf 207}, 511

\noindent 
Haehnelt, M. G., Rees, M. J. 1993, {\it MNRAS}~ {\bf 263}, 168

\noindent 
Hernquist, L., Ostriker, J. P., {\it ApJ}~ {\bf 386}, 362

\noindent
Hockney, R., W., Eastwood, J. W., 1981. { `Computer
Simulations Using Particles'}, McGraw-Hill

\noindent
Hunter, C, Qian, E. 1993, {\it MNRAS} {\bf 262}, 401

\noindent
Jaffe, W., Ford, H. C., O'Connell, R. W., van den Bosch, F. C.,
Ferrarese, L.  1994, {\it AJ}~ {\bf 108}, 1567

\noindent
Joseph, C. L., Merritt, D., Olling, R., Valluri, M. {\it et al.} 2000 {\it in preparation}

\noindent
Kawai, A., Fukushige, T., Taiji, M., Makino, J., Sugimoto, D. 1997,
{\it PASP}~ {\bf 49}, 607

\noindent
Kormendy, J. 1988.  In { `Supermassive Black Holes'}, ed. M. Kafatos
(Cambridge: Cambridge Univ. Press)

\noindent
Kormendy, J., Bender, R., Evans, A. S., Richstone, D. 1998, {\it AJ} {\bf
115}, 1823

\noindent
Kormendy, J., Richstone, D. 1995, {\it ARA\&A} {\bf 33}, 581

\noindent
Lauer, T. {\it et al.} 1995, {\it AJ} {\bf 110}, 2622

\noindent
Lee, M. H., Goodman, J. 1989, {\it ApJ} {\bf 343}, 594 (LG89)

\noindent
Leeuwin, F., Combes, F., Binney, J. 1993, {\it MNRAS} {\bf 262}, 1013

\noindent
Leeuwin, F., Dejonghe, H. 1998. In  `Galaxy Dynamics',
ed. D. Merritt, M. Valluri \& J. Sellwood  (ASP Conf. Series) 

\noindent
Lynden-Bell, D., 1962, {\it MNRAS} {\bf 123}, 457

\noindent
Lynden-Bell, D., 1967, {\it MNRAS} {\bf 136}, 101

\noindent
Magorrian, J., Tremaine, S. 1999, {\it MNRAS} {\bf 309}, 447

\noindent
Makino, J., Ebisuzaki, T. 1996, {\it ApJ} {\bf 465}, 527

\noindent
Makino, J., Taiji, M., Ebisuzaki, T., Sugimoto, D. 1997, {\it ApJ} {\bf
480}, 432

\noindent Merritt, D. 1998. In  `Galaxy Dynamics',
ed. D. Merritt, M. Valluri \& J. Sellwood  (ASP Conf. Series) 

\noindent
Merritt, D., Quinlan, G. 1998, {\it ApJ} {\bf 498}, 625

\noindent
Nakano, T., Makino, J. 1999, {\it ApJ} {\bf 510}, 155

\noindent
Nieto, J.-L., Bender, R. 1989, {\it A\&A} {\bf 215}, 266 

\noindent
Norman, C.A., May, A., van Albada, T. S. 1985, {\it ApJ} {\bf 296}, 20

\noindent
Peebles, P. J. E. 1972, {\it Gen. Rel. Grav.} {\bf 3}, 61

\noindent
Press, W. H., Flannery, B. P., Teukolsky, S. A., Vetterling, W. T.,
1986. `Numerical Recipes', Cambridge Univ. Press

\noindent
Quinlan, G. D., Hernquist, L. 1997, {\it New Ast.} {\bf 2}, 533 

\noindent
Quinlan, G. D., Hernquist, L., Sigurdsson, S. 1995, {\it ApJ} {\bf 440}, 554

\noindent
Richstone, D. {\it et al.}  1998, {\it Nature} {\bf 395}, 14

\noindent 
Saha, P., Tremaine, S. 1992, {\it AJ} {\bf 104}, 1633

\noindent
Scorza, C., van den Bosch, F. C. 1998, {\it MNRAS} {\bf 300}, 469

\noindent
Sigurdsson, S., Hernquist, L., Quinlan, G. D. 1995, {\it ApJ} {\bf 446}, 75-85.

\noindent 
Stiavelli, M. 1998, {\it ApJ} {\bf 495}, 91

\noindent
Tremaine, S. 1997. In `Unsolved Problems in Astrophysics',
ed. J. Ostriker (Princeton)

\noindent 
Tremaine, S. {\it et al.} 1994, {\it AJ} {\bf 107}, 634

\noindent
Udry, S. 1993, {\it A\&A} {\bf 268}, 35


\noindent
van den Bosch, F. C., Ferrarese, L., Jaffe, W., Ford, H., O'Connell,
R. W. 1994, {\it AJ} {\bf 108}, 1579

\noindent
van der Marel, R. P. 1999, {\it AJ} {\bf 117}, 744

\noindent 
van der Marel, R. P. 1997. In `Galaxy Interactions at Low and High
Redshift', IAU Symp. 186, eds. D. Sanders \& J. Barnes (Kluwer)

\noindent
van der Marel, R., de Zeeuw, T., Rix, H.-W. 1997, {\it ApJ} {\bf 488}, 702

\noindent 
Wachlin, F. C., Rybicki, G. B., Muzzio, J. C. 1993, {\it MNRAS}
{\bf 262}, 1007

\noindent
Wisdom, J., Holman, M. 1991, {\it AJ} {\bf 102}, 1528

\noindent
Young, P. 1980, {\it ApJ} {\bf 242}, 1232

\label{lastpage}

\end{document}